\documentclass[acmtog,authorversion,nonacm]{acmart} 
\usepackage{subcaption}
\usepackage{rotating}
\usepackage{bm} % \bm for bold italics
\usepackage{wrapfig} % for that one inset figure
\AtBeginDocument{%
  \providecommand\BibTeX{{%
    \normalfont B\kern-0.5em{\scshape i\kern-0.25em b}\kern-0.8em\TeX}}}

\newcommand{\figref}[1]{Fig.~\ref{#1}}
\newcommand{\myeqref}[1]{Eq.~\ref{#1}}
\newcommand{\tabref}[1]{Table~\ref{#1}}
\newcommand{\secref}[1]{Section~\ref{#1}}
\def\R{\mathbb{R}}

\usepackage{graphicx}
\begin{document}
\title{Unphased Wrinkles: Estimating cloth elasticity parameters using a frequency-based loss}

%%
%% end of the preamble, start of the body of the document source.

\author{Egor Larionov}
\affiliation{
  \institution{Meta Reality Labs Research}
  \country{USA}
}
\author{Marie-Lena Eckert}\affiliation{
  \institution{Meta Reality Labs Research}
  \country{Switzerland}
}
\author{Katja Wolff}\affiliation{
  \institution{Meta Reality Labs Research}
  \country{Switzerland}
}
\author{Tuur Stuyck}\affiliation{
  \institution{Meta Reality Labs Research}
  \country{USA}
}

% \teaser{
\begin{teaserfigure}
\centering % used for centering Figure
\includegraphics[width=\textwidth]{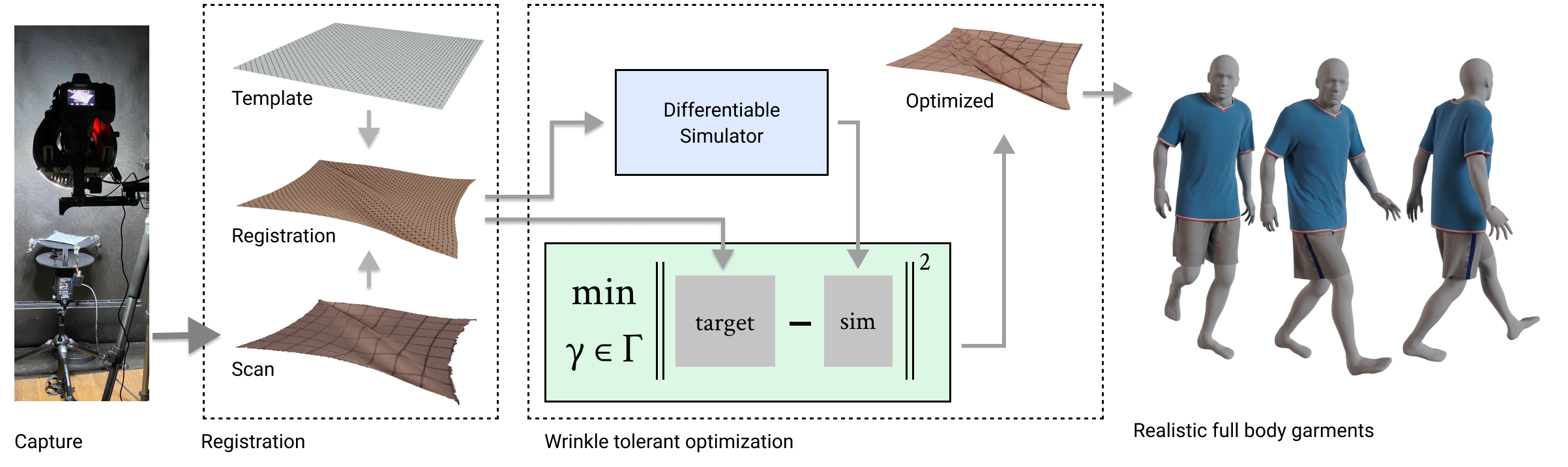}
\caption{\label{fig:pipeline} \emph{Parameter Estimation.} With our method, we decouple the cloth capture (left) from the parameter optimization (middle-right) using NR-ICP mesh registration (middle-left). The optimization pass is able to handle wrinkled cloth, which greatly simplifies the capture process. With the optimized parameters $\gamma$, we generate realistic full-body cloth simulation (right) bypassing laborious manual parameter tuning.}
% \label{fig:teaser} % label to refer figure in text
\end{teaserfigure}

\renewcommand{\shortauthors}{Larionov et al.}

\begin{abstract}
% Clothing plays a vital role in real life and hence, is also important for virtual realities and virtual applications, such as online retail, virtual try-on, and real-time digital avatar interactions. However, choosing the correct parameters to generate realistic clothing requires expert knowledge and is often an arduous manual process.
Generating realistic clothing for virtual applications like online retail and digital avatars is crucial but requires expert knowledge of 3D tools to generating believable simulations.  Recently, a number of works proposed to estimate cloth material properties from specialized capture setups. However, these systems tend to be monolithic, complex and expensive. We propose a simplified method for automatically determining parameters based on easily captured real-world fabrics.  While existing methods carefully design experiments to isolate stretch parameters from bending modes, we embrace that stretching fabrics causes wrinkling and propose a novel specialized loss for comparing wrinkled fabrics.  We designed our objective function to capture material-specific behavior, resulting in similar values for different wrinkle configurations of the same material. We estimate bending first, given that membrane stiffness has little effect on bending. We use differentiable simulation to find an optimal set of parameters that minimizes the difference between simulated cloth and deformed target cloth. Furthermore, our pipeline decouples the capture method from the optimization by registering a template mesh to the scanned data. These choices simplify the capture system and allow for wrinkles in scanned fabrics. %We use a differentiable cloth simulator, which is capable of real-time simulation.
We demonstrate our method on captured data of three different real-world fabrics and on three digital fabrics produced by a third-party simulator.
\end{abstract}
% TODO: fill this out


% \printccsdesc


\maketitle

\section{Introduction}

Clothing plays an important cultural role and enables a form of self-ex\-pres\-sion. %The fashion industry has a total worth of roughly 1.5 trillion US dollars that is even expected to double by 2025.
Traditional fashion houses are becoming more and more aware of the virtual landscape and are experimenting with virtual collections in the Metaverse in addition to real-world fashion lines and virtual try-on. Virtual clothing research has reduced production costs and time to market. To reproduce familiar experiences and opportunities for self-expression in the virtual world, it is essential to be able to simulate a large variety of different types of clothing made from vastly different materials and structures. Unfortunately, creating realistic clothed human animations currently involves expert knowledge and hours of tweaking simulation parameters to obtain the desired clothing look.

With this project, we aim to reduce the time and cost required to capture and optimize for parameters to create realistic cloth simulations of real-world materials. 
While much of the effort in simulation research targets improving numerical accuracy and minimizing computation time, we instead focus on improving the realism of an existing cloth simulation method.

\begin{figure}
    \centering
    \includegraphics[width=0.49\linewidth,trim=0 300 0 300,clip]{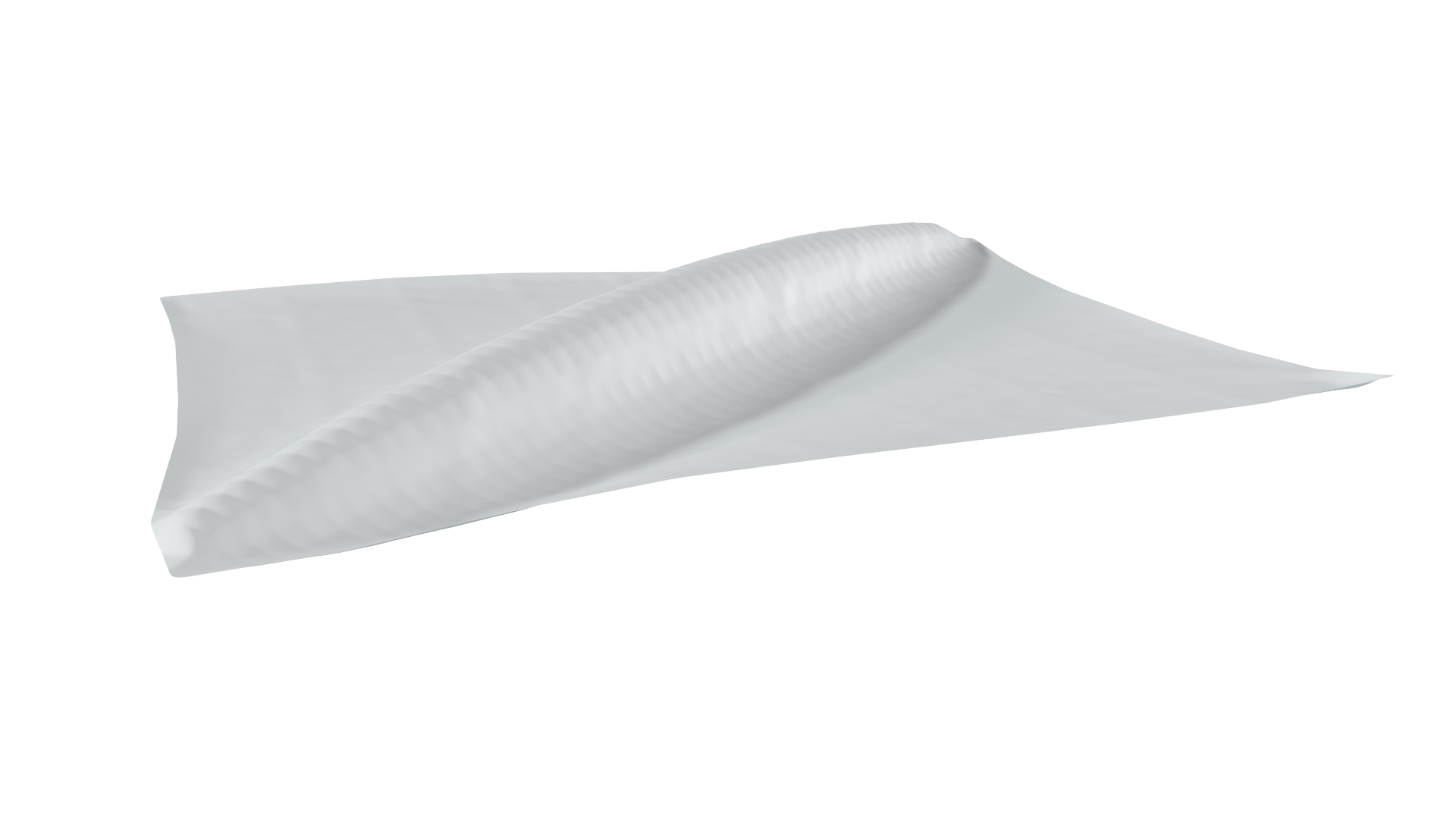}
    \includegraphics[width=0.49\linewidth,trim=0 300 0 300,clip]{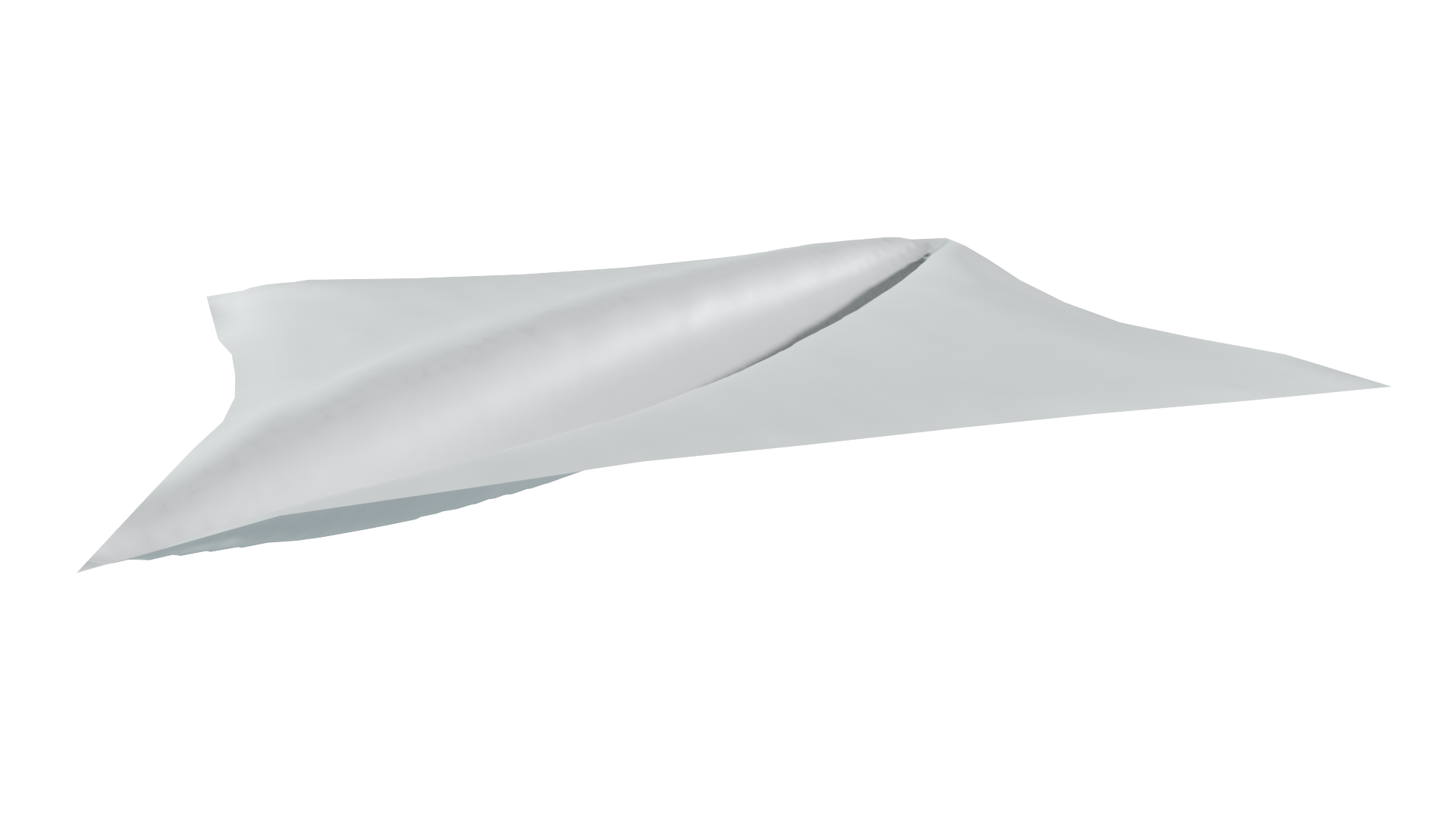}
    \includegraphics[width=0.49\linewidth,trim=0 400 0 400,clip]{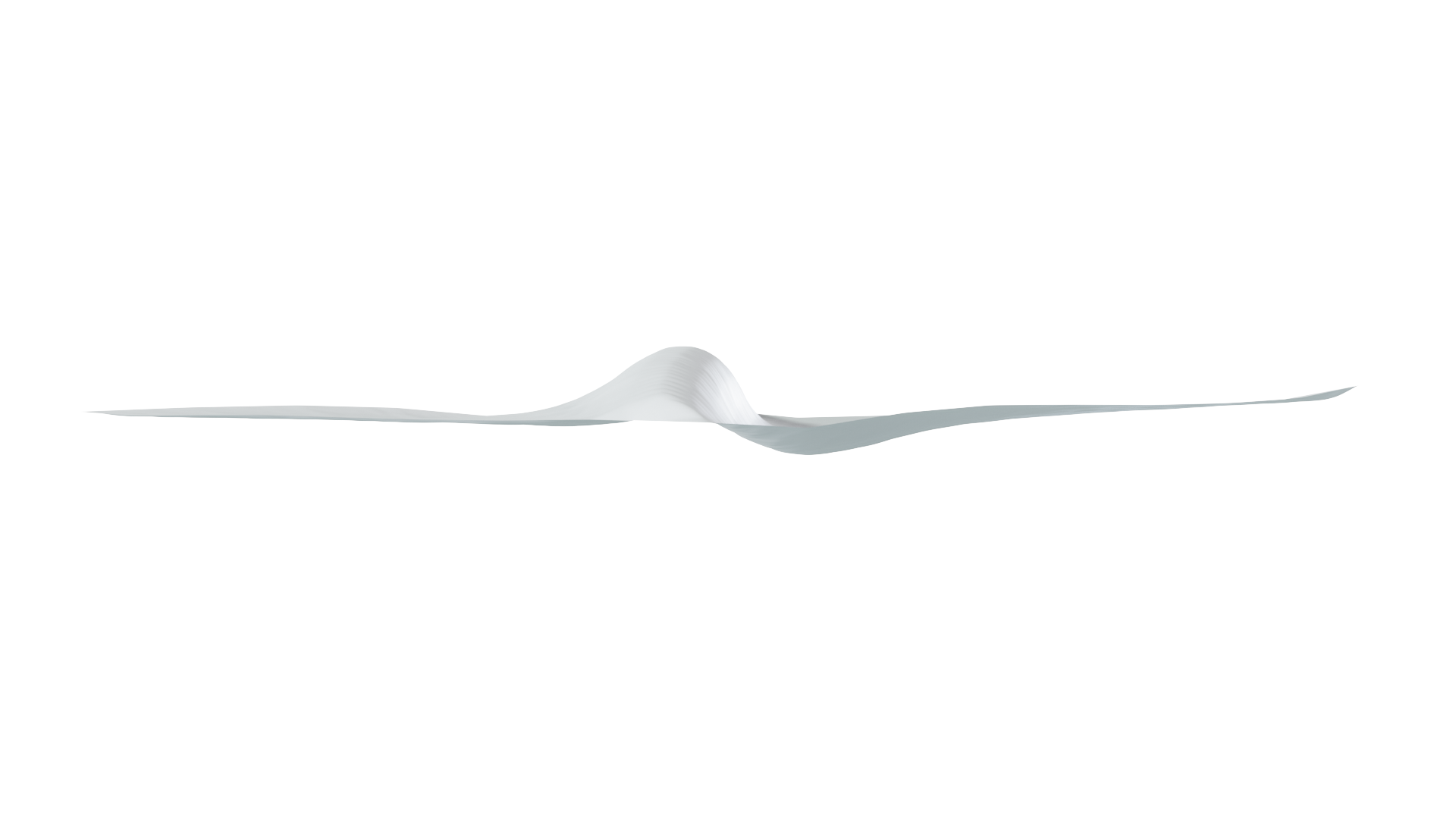}
    \includegraphics[width=0.49\linewidth,trim=0 400 0 400,clip]{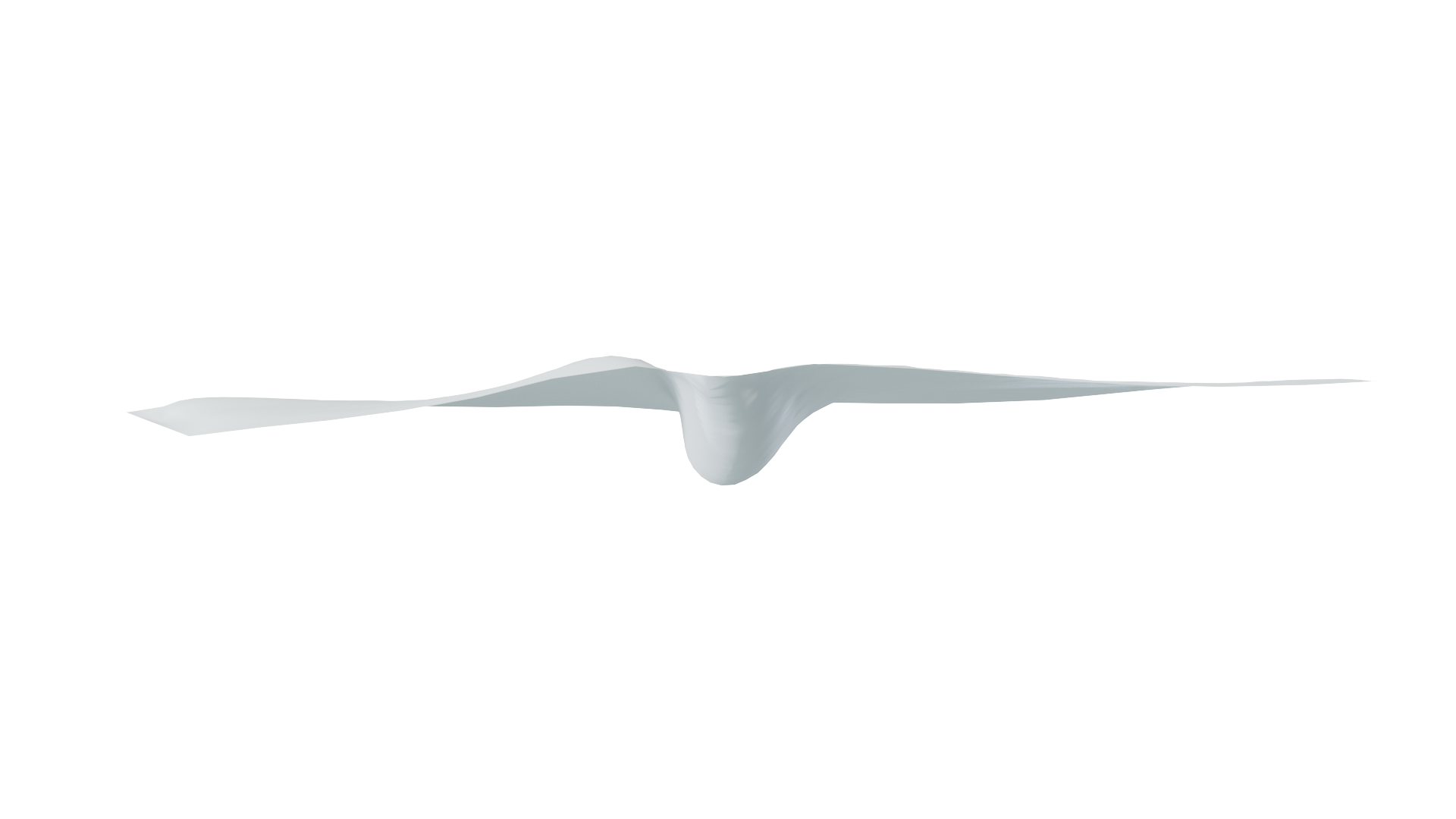}
    \caption{\emph{Wrinkle bifurcation}: two valid cloth configurations with contrary wrinkle patterns shown from above (top) and side (bottom). The same cloth swatch is pulled with the same force at opposite corners.}
    \label{fig:wrinkle}
\end{figure}

Consider a square piece of fabric being pulled at each of the four corners as depicted in~\figref{fig:wrinkle}, which shows a 3D reconstruction of a piece of real fabric. By slightly readjusting the same fabric, we can produce two different configurations shown on the left and right for the Figure. Naively comparing the vertex positions of the resulting meshes would yield a nontrivial difference in spite of them corresponding to the same piece of fabric. This poses a real challenge for estimating material parameters from such captures.

To that end, we present a material estimation method, which leverages a frequency-based loss to robustly deal with cloth wrinkling.
This enables us to process easy to reproduce cloth captures. In general, capturing cloth samples is often difficult to control due to environmental, material shape memory, and hysteresis effects. Small perturbations can lead to significantly different cloth equilibrium states, which typically makes the optimization heavily biased towards the specific captured sample of the fabric. To overcome these limitations we introduce the following contributions:
\begin{itemize}
    \item A novel robust objective function that operates in frequency space and is able to capture material-specific behavior, independent of the current cloth wrinkle state.
    \item A simplified pipeline for estimating material elasticity properties of cloth that is decoupled from the real-world capture system through a template registration process.
    \item A real-world validation of XPBD cloth simulation with numerical comparisons of simulated results and captured scans.
    \item A set of material parameters for the standard St. Venant-Kirchhoff (StVK) cloth material model for three distinct common materials.
\end{itemize}

\section{Related Work}

Given the high relevance of virtual textile simulation both in research and manufacturing, material estimation has been a topic of investigation in the research community \cite{luible2008study} and industry \cite{clo, browz, optitex}. We highlight relevant work.

\textbf{Material parameter estimation} has been a focus point in computer graphics. While some techniques address the full scope of dynamic simulation parameters using video sequences \cite{bhat}, we focus instead on elasticity, which can be inferred from static captures of deformed cloth.
Wang~et~al.~\shortcite{wang2011data} have proposed a non-standard piece-wise material model and fit 39 parameters to a set of sparse correspondences. While sophisticated, their scheme can exhibit unlikely Poisson's ratios below zero or above one. Runia~et~al.~\shortcite{runia2020cloth} proposes an embedding function which maps physically similar states to nearby points which contributes to learning cloth material parameters from videos.   
Prior work focusing on mass-spring and diagonalized StVK models \cite{miguel12} produce promising parameter estimations, but at the cost of requiring a complex and expensive capture setup. Furthermore, their method omits the Poisson effect, which greatly limits the applicability of their results in modern cloth simulators. Both works employ nodal positions to estimate approximation error, which makes their methods sensitive to wrinkles in captured data.   
% Prior work focusing on on mass-spring models and and diagonalized StVK material model \cite{miguel12} produce satisfactory results but at the cost of requiring a complex and expensive setup. The method omits Poisson effect and employs node positions and boundary forces to guide the optimization. \cite{clyde17} Complex fitting procedure (specialized objectives). Uses ASTM measurements (angles and forces).}
Clyde~et~al.~\shortcite{clyde17} propose a more sophisticated nonlinear material model to better fit existing standard ASTM measurements in large deformation scenarios. In contrast, we focus on improving the optimization method to handle data coming from non-standard and potentially cheaper setups.
% In the limited setting of video observations of the fabric, the material parameters can be estimated using a perceptually motivated metric \cite{bhat}.
A different line of work attempts to estimate internal friction in cloth \cite{miguel2013modeling}. Learning-based methods have been used for the purpose of material prediction from real images while being trained on synthetic data \cite{rodriguez2023will} and to predict bending stiffness parameters directly from drape features captured in the real world \cite{feng2022learning}. Other methods estimate simulation parameters using neural networks from a static drape \cite{ju2020estimating, TagInfoCloth}, images~\cite{dominguez2024practical} or from videos \cite{yang2017learning}.

\textbf{3D shape descriptors} provide a compact encoding of certain properties of 3D objects and can be used for classification, registration or correspondence matching tasks \cite{ovsjanikov2012functional, litany2017deep}. Shape descriptors have successfully been developed leveraging spherical harmonics \cite{saupe20013d, vranic2001tools} as well as learned features \cite{fang20153d}. Lopez~et~al.~\cite{lara2017comparative} provide an extensive comparative analysis of shape descriptors for 3D objects. 

\textbf{Differentiable simulation \& system identification} has been researched \cite{liang2019differentiable, qiao2020scalable} to infer material parameters from observations \cite{gradsim21, strecke2021_diffsdfsim, hu2019chainqueen, du2021diffpd, stuyck2023diffxpbd}. Differentiable simulation has been successfully used to control contact-rich cloth simulation \cite{li2022diffcloth} as well as to recover simulation-ready assets from real data \cite{li2023diffavatar, Wang_2023_CVPR}. A Bayesian differentiable cloth framework to estimate cloth materials from limited data is presented by Gong~et~al.~\shortcite{Gong_Bayesian_2024}. Another topic of focus has been to bring real deformable objects into the virtual world using sparse observations and interaction of the object with its environment \cite{weiss2020correspondence, veo}.

\textbf{Cloth simulation} has been a topic of research in computer graphics for many decades. Starting from the earlier work \cite{baraff1998large}, many improvements have been proposed to increase accuracy, stability and performance \cite{muller2007position, liu2013fast, bouaziz2014projective, macklin2016xpbd, overby2017admm, stuyck2018cloth}. 

In summary, the research field has made great progress while several issues still remain. Deeply intertwined pipelines of closely coupled capture and optimization systems complicate the combination of different techniques. Furthermore, multiple prior works use complex capture setups that are difficult to scale. Difficulties with wrinkling modes has yet to be addressed. Some methods \cite{TagInfoCloth} focus on careful draping hardware, while others develop a drape specific metric \cite{Gong_Bayesian_2024} to bypass this problem. Moreover, many techniques are difficult to generalize to other solvers, because they employ non-standard material models \cite{wang2011data, clyde17}. %In this work, we leverage position-based simulation of compliant dynamics, often referred to as \emph{XPBD} \cite{macklin2016xpbd}, see Section \ref{sec:simulation}. 

\section{Method}

% \begin{figure*}[tb]
%     \centering
%     \includegraphics[width=1.0\textwidth]{images/pipeline5.PNG}
%     \caption{\label{fig:pipeline} Our pipeline decouples the capture (left) from the optimization (right) using NR-ICP mesh registration (middle). }
%     %\MLE{1. enlarge cloth images for registration and increase contrast if possible (same background as block would be nice but that's prob. too much work) and label each image (template mesh, input mesh, registered mesh (or omit mesh for each)), 2. create block around "Diff cloth sim" and equation block, 3. if not too much work, center the captions for each sub-block ("Capture", "Registration", "Material..") under the blocks}
% \end{figure*}

Our goal is to estimate the material properties needed for the cloth simulator to represent particular cloth materials as realistically as possible.
To achieve this, we develop a pipeline with three independent stages made from four components as visualized in \figref{fig:pipeline}. Initially, a \emph{capture system} gathers point or mesh data of a deformed rectangular cloth swatch in various configurations. Then, a template mesh is \emph{registered} to the captured data using a landmark registration technique: the non-rigid iterative closest point (NR-ICP). Finally, the registered mesh is compared to a \emph{simulation} of the template mesh, and an optimal set of parameters is found to align both as closely as possible by using \emph{least squares optimization}.
%Finally, this mesh is compared to a \emph{simulation} of the \MLE{template ?} same mesh, and by using \emph{least squares optimization}, an optimal set of parameters is picked to generate the closest simulation possible to the target registered mesh.

%\MLE{create and use new command myreffig to ensure consistency when referencing a Figure, e.g. Figure 1 or Fig. 1, same for table etc.}.
% EL: Using new commands \figref and \tabref (to be consistent with \eqref)

\subsection{Capture System}

\begin{figure}
    \centering
    \includegraphics[width=0.48\textwidth]{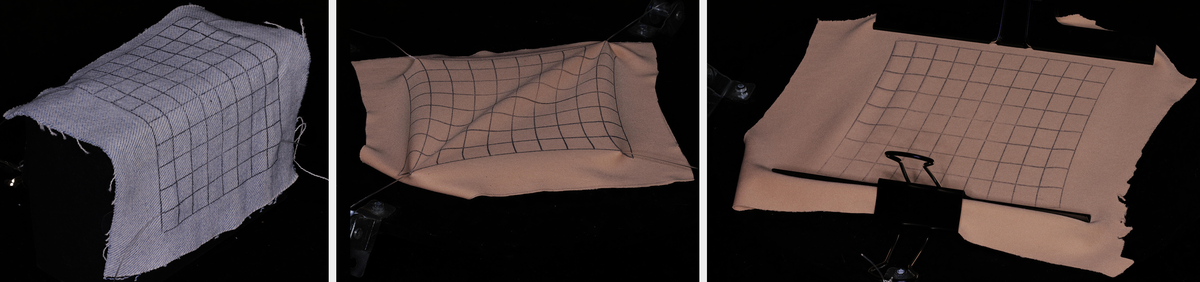}
    \caption{\emph{Cloth capture.} Our simple cloth capture system records the cloth swatch under different force applications to drive the optimization.}
    \label{fig:captureSystem}
\end{figure}

We propose a lightweight cloth capture setup that can be easily reproduced with minimal cost. Cloth swatches are suspended using clamps that apply a controllable external force through weights, see~\figref{fig:captureSystem}. The images are processed using \emph{Agisoft Metashape} \cite{agisoft} to produce a high-resolution texture mapped 3D mesh. A free and open-source alternative is Meshroom \cite{alicevision2021}.
The swatch is stamped with a regular grid pattern to provide landmarks for the registration process described below.

\subsection{Template Registration} \label{sec:registration}

\begin{figure}
    \centering
    \includegraphics[width=0.4\textwidth]{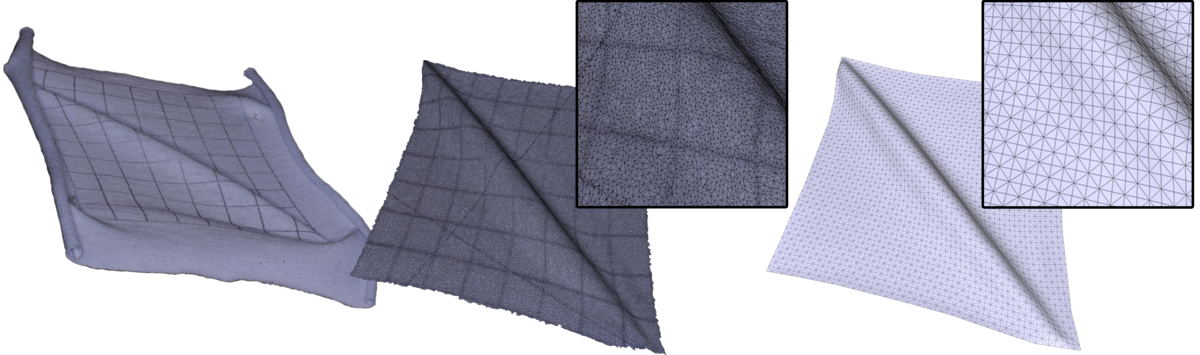}
    \caption{\emph{Registration.} Capture setup (left), scanned mesh (middle), and registered mesh (right) with magnified regions in the insets.\vspace{-1.5em}
}
    \label{fig:registration}
\end{figure}
In contrast to prior research, our simulation targets are decoupled from the captured scans through mesh registration, see~\figref{fig:registration}. We create a triangulated regular grid mesh representing the simulated cloth swatch. This template is then registered to the scan by pairing vertices of the grid with landmarks on the scanned mesh via the NR-ICP method~\cite{li2008global}. Vertices that are not aligned with the stamped grid pattern are projected along the mesh normal towards the surface of the scanned mesh to improve the position estimate in the normal direction.
While more accurate methods have been developed for establishing dense correspondences between meshes, we stress that any stage of our pipeline -- including template registration -- is loosely coupled and can be replaced independently.

\subsection{Simulation}
\label{sec:simulation}

%Cloth simulation in computer graphics has been a topic of research for many decades. Starting from the earlier work \cite{baraff1998large}, many improvements have been proposed to increase accuracy, stability and performance \cite{muller2007position, liu2013fast, bouaziz2014projective, macklin2016xpbd, overby2017admm, stuyck2018cloth}. In this work, we leverage position-based simulation of compliant dynamics, often referred to as \emph{XPBD}. 

%\subsubsection{Extended Position Based Dynamics}

% why did we use xpbd and not something else? 
We leverage position-based simulation of compliant dynamics (XPBD) \cite{macklin2016xpbd} as it often obtains much better performance compared to other solvers. It uses an iterative Gauss-Seidel solution for the linearized equations of motion. The method can be easily parallelized \cite{fratarcangeli2016vivace}, enabling interactive or real-time simulations on common modern hardware. Note that other simulation methods are compatible with our proposed capture and optimization method including our proposed frequency-based loss. 
%XPBD \cite{macklin2016xpbd} is a recent constraint-based simulation model that often obtains much better performance compared to expensive non-linear solvers. It uses an iterative Gauss-Seidel solution for the linearized equations of motion. The method can be easily parallelized \cite{fratarcangeli2016vivace} and implemented on hardware such as multi-core CPUs and GPUs, enabling interactive or real-time simulations on common modern hardware.

The method aims to solve Newton's equations of motion
\begin{align}
    \mathbf{M}\ddot{\mathbf{x}} = -\nabla U(\mathbf{x}),
\end{align}
where $\mathbf{x} \in \R^{3n}$
% \KW{not n x 3?}
encodes $n$ vertex positions (of the cloth mesh in this case) and $\mathbf{M}$ is the mass matrix computed from element volumes and constant material density $\rho$.
% \KW{Maybe a dumb question: the density is not part of the optimized parameters, right? How is it chosen?}
The energy potential $U(\mathbf{x})$ needs to be specified in terms of a vector of constraint functions $\mathbf{C} = \left[C_1(x), C_2(x), \cdot \cdot \cdot, C_m(x)\right]^{\top}$ as
\begin{equation}
U(\mathbf{x}) = \frac 12 \mathbf{C}(\mathbf{x})^{\top} \bm{\alpha}^{-1} \mathbf{C}(\mathbf{x}),
\label{eq:xpbdPotentialEnergy}
\end{equation}
where \(\bm{\alpha}\) is a block diagonal compliance matrix. %Any energy that can be written this way is suitable for XPBD.
%Using implicit Euler time integration, the algorithm reduces to solving for the constraint multiplier updates $\Delta \bm{\lambda}$ with %\MLE{should we specify what the tilde means at $\tilde{\bm{\alpha}}$? Reviewers might criticize if we don't}
The method first solves for the constraint multiplier updates $\Delta \bm{\lambda}$ with
\begin{align}
    (\nabla \mathbf{C}(\mathbf{x}_i)^{\top}\mathbf{M}^{-1}\nabla\mathbf{C}(\mathbf{x}_i) + \tilde{\bm{\alpha}}) \Delta \bm{\lambda} = - \mathbf{C}(\mathbf{x}_i) - \tilde{\bm{\alpha}}\bm{\lambda}_i, \label{eq:xpbd}
\end{align}
where $\tilde{\bm{\alpha}} = \frac{\bm{\alpha}}{\Delta t^2}$, followed by a position update
\begin{align}
    \Delta \mathbf{x} = \mathbf{M}^{-1} \nabla \mathbf{C}(\mathbf{x}_i)\Delta \bm{\lambda}.
    \label{eq:posUpdate}
\end{align}

%The system in~\myeqref{eq:xpbd} is typically solved using Gauss-Seidel- or Jacobi-style updates.

\subsubsection{Elasticity Model}
\label{sec:stvk}

To demonstrate our method, we employ an orthotropic StVK membrane energy model along with simple discrete bending \cite{bender17} for modelling cloth. Note that other models can be employed as long as they are differentiable. The chosen model suggests a per-element inverse compliance matrix of the form
$$
	\bm{\alpha}_{\triangle}^{-1} = A\begin{bmatrix} C_{00} & C_{01} & \\ C_{01} & C_{11} & \\ & & C_{22} \end{bmatrix},
$$
where \(A\) is the area of a single triangle and \(C_{ij}\) are the compliance coefficients. The constraint function for each triangle is then defined to be the Green strain $\epsilon$ in Voigt notation
\begin{equation}
	\mathbf{C}_{\triangle}(\mathbf{x}) = (\epsilon_{uu}, \epsilon_{vv}, 2\epsilon_{uv})^{\top},
\end{equation}
where subscripts $u$ and $v$ indicate warp and weft directions, respectively.
With orthotropic Young's moduli $E_u, E_v$ for modelling distinct warp and weft behavior of the fabric, Poisson's ratios $\nu_{uv}, \nu_{vu}$ and shear modulus $\mu$, we have
\begin{equation}
\begin{bmatrix} C_{00}\!\!\! & C_{01}\!\!\! &\!\!\! \\ C_{01}\!\!\! & C_{11}\!\!\! & \!\!\!\\ \!\!\!& \!\!\!& \!\!\!C_{22} \end{bmatrix} = 
\frac{1}{1-\nu_{uv}\nu_{vu}}\begin{bmatrix} E_u\!\!\! & \nu_{vu}E_u\!\!\! &\!\!\! \\ \nu_{uv}E_v \!\!\!& E_v \!\!\!& \!\!\!\\ \!\!\!& \!\!\!& \!\!\!\mu(1-\nu_{uv}\nu_{vu}) \end{bmatrix}.
\label{eq:paramsCorrespondence}
\end{equation}
Note that this matrix is symmetric since \(\nu_{vu}E_u = \nu_{uv}E_v\) and the Poisson's ratio \(\nu_{vu}\) corresponds to a contraction in direction \(u\) when an extension is applied in direction \(v\).  In the following sections, we abbreviate $\nu_{uv} = \nu$ whereas $\nu_{vu}$ is computed from $\nu, E_u,$ and $E_v$.
% By taking the inverse, we can also write the matrix \(\bm{\alpha}\) as
% $$
% \bm{\alpha} = 
% \frac{1}{A}\begin{bmatrix} \frac{1}{E_u} & -\frac{\nu_{vu}}{E_v} & \\ -\frac{\nu_{uv}}{E_u} & \frac{1}{E_v} & \\ & & \frac{1}{\mu} \end{bmatrix}
% $$

The bending constraint \cite{bender17} is defined for each pair of adjacent triangles ($\mathbf{x}_1$, $\mathbf{x}_3$, $\mathbf{x}_2$), ($\mathbf{x}_1$,$\mathbf{x}_2$,$\mathbf{x}_4$) as the angle strain
\begin{align*}
    \def\facenmla{\mathbf{x}_{2,1}\times\mathbf{x}_{3,1}}
    \def\facenmlb{\mathbf{x}_{2,1}\times\mathbf{x}_{4,1}}
    C_{\text{bend}} = \arccos\left(\frac{\facenmla}{\|\facenmla\|} \cdot \frac{\facenmlb}{\|\facenmlb\|} \right) - \phi_0,
\end{align*}
where $\phi_0$ is the rest dihedral angle and $\mathbf{x}_{i,j} = \mathbf{x}_i - \mathbf{x}_j$ are edge vectors between vertices $i$ and $j$. The inverse compliance matrix is then given by the scalar bending stiffness: $\bm{\alpha}_{\text{bend}}^{-1} = [b]$.

\subsubsection{Differentiable Simulation}

To enable gradient-based optimization, we require the simulation framework to be able to compute derivatives of the simulated positions with respect to the material parameters $\gamma$. Given that both the bending energy and the StVK membrane energy are differentiable, it suffices to compute the derivative of the position update in \myeqref{eq:posUpdate}, which involves differentiating through \myeqref{eq:xpbd}.
In contrast to work like DiffXPBD \cite{stuyck2023diffxpbd} or DiffPD \cite{du2021diffpd}, which use the adjoint method to obtain differentiable formulations of the simulation model, we propose to use forward differentiation of the required quantities. Our choice of differentiation results in a much more efficient pipeline, which is able to compute all required quantities in a single forward pass. The derivatives are computed analytically and accumulated in tandem with the Gauss-Seidel or Jacobi iterates. This is in contrast to adjoint-based methods which require an additional costly backward pass. 

\subsection{Estimating Elastic Material Parameters}

In this section, we describe our method for finding a set of parameters \(\gamma\) required to reproduce a cloth shape captured under the influence of external forces. We focus on estimating static elasticity parameters that define the stress-strain relationship. Estimating parameters that describe how cloth behaves during motion, i.e., friction or damping, remains an interesting avenue for future work. 

\subsubsection{Optimization Problem} \label{sec:optiProblem}
As output from the capture pipeline, we obtain a cloth mesh in a deformed configuration at rest. We introduce \emph{shape descriptors} \(\mathbf{s}\) that represent the cloth's shape in different ways, see Sec.~\ref{sec:shapedescriptors}. Below,
\(\mathbf{s}_{\text{target}}\) is used to describe the target shape. To obtain more information about the stress-strain relationship in the material and to resolve global scaling ambiguities, we additionally use force data from some of the boundary nodes. While other works have used full force vectors, we opt to capture only force magnitudes \(\mathbf{f}_{\text{target}}\) to avoid the need for complex and costly setups necessary to collect accurate directional forces. We then match the corresponding shape $\mathbf{s}_{\text{sim}}$ and boundary forces $\mathbf{f}_{\text{sim}}$ of a simulation to the target by finding suitable material properties. In general, we can stack as many meshes as needed into $\mathbf{s}$ and $\mathbf{f}$.

We optimize directly over the compliance coefficients and bending parameter by choosing the parameter set to be
\begin{equation}
    \gamma := (C_{00}, C_{11}, C_{01}, C_{11}, b), \ \gamma\in\Gamma,
    \label{eq:params}
\end{equation}
where $\Gamma$ is a rectangular constraint set of feasible material parameter combinations.
The chosen parameter set \(\gamma\) is a better candidate for optimization compared to Young's moduli and Poisson's ratio due to its good relative scaling.
%\MLE{what kind of adjustment? a bit unclear}
% Furthermore, the rectangular constraints eliminate many infeasible parameter combinations beforehand. 
To penalize unrealistic Poisson's ratios $\nu$ above one, we add the following penalty
%\MLE{we should add a term for the penalty; would be even nicer to directly add it to Eq. 5 to make it very clear}
$$
W_\nu = \max(0, \nu - 1).
$$
Putting everything together forms our final optimization problem
\begin{align}
	\min_{\gamma\in \Gamma}  \|\mathbf{r}_{\text{sim}}(\gamma) - \mathbf{r}_{\text{target}}\|^2 + s_\nu W_\nu, \label{eq:origOpt}
\end{align}
where \(\mathbf{r}^{\top} = (\mathbf{s}^{\top}, \mathbf{f}^{\top})\) is the vector of stacked shape descriptors $\mathbf{s}$ and boundary force magnitudes $\mathbf{f}$. 
Here, \(\mathbf{r}_{\text{sim}}\) corresponds to a simulation after \(N\) time steps of a quasi-static solve initialized to the target vertex positions, whereas $\mathbf{r}_{\text{target}}$ are the shape descriptors and measured force magnitudes of the registered mesh as described in \secref{sec:registration}. We weigh the Poisson penalty high with $s_\nu = 1e8$.
To solve \myeqref{eq:origOpt}, we employ the Ceres nonlinear least squares solver \cite{ceres-solver}.

\subsubsection{Shape Descriptors} \label{sec:shapedescriptors}
Due to the bifurcation behavior of wrinkling, deformed cloth can reach distinct equilibrium states depending on the deformation trajectory, initial conditions, and material shape memory effects. Many prior methods rely on precise control over cloth swatches to measure specific parameters \cite{wang2011data, miguel12, clyde17, TagInfoCloth}. In particular, it is convenient to produce deformations that are affected by just one or a few material parameters. For instance, the bending angle in a piece of cloth, as it hangs from the edge of a table, is determined by the bending parameter while being largely insensitive to changes in other stiffness parameters. Unfortunately, it is difficult to design experiments that isolate material properties like shear stiffness and Poisson's ratio, since cloth tends to buckle in such scenarios producing wrinkles that inherently couple bending with other parameters.

Due to the complicated cloth behavior, minimizing a direct comparison of 3D vertex positions would not produce robust results. 
%As outlined above, the evaluation of the objective in \myeqref{eq:origOpt} results in a very different value
% \KW{very different cloth parameters $\gamma$?}
%for both equilibrium states in \figref{fig:wrinkle} when choosing \(\mathbf{s}\) to be a stacked locator of vertex positions $\mathbf{p}$, although both target swatches are generated with the same material properties. 
In general, a positional loss encoding is sensitive to small differences in the captured cloth. Thus, we developed a novel objective function that reports a similar error for wrinkle patterns generated by the same material. As a result, we are able to process coupled-parameter captures, demonstrating robustness to different configurations due to different initial conditions. This in turn, allows for simpler and more affordable capture pipelines. We evaluate the following shape descriptors:
\begin{equation}
\begin{aligned}
\mathbf{s}_{\text{pos}}(\mathbf{p}) &= \mathbf{p}, \\
\mathbf{s}_{\text{energy}}(\mathbf{p}) &= \mathbf{U}_{\text{triangle}}(\mathbf{p}) + \mathbf{U}_{\text{bend}}(\mathbf{p}), \\ 
\mathbf{s}_{\text{strain}}(\mathbf{p}) &= \mathbf{C}_{\text{triangle}}(\mathbf{p}) + \mathbf{C}_{\text{bend}}(\mathbf{p}), \\
\mathbf{s}_{\text{freq}}(\mathbf{p}) &= |\mathtt{FFT}(\mathbf{p})|,
\label{eq:shapeDescriptors}
\end{aligned}
\end{equation}
where \(\mathbf{U}\) is either a triangle or bending \emph{energy} and \(\mathbf{C}\) is the \emph{strain}
%\MLE{strain isn't defined in that section afaik}
as defined in \secref{sec:stvk}.
The position descriptor $\mathbf{s}_{\text{pos}}$ represents the swatch shape in terms of vertex positions, leading to evaluating the absolute position of wrinkles. In contrast, $\mathbf{s}_{\text{energy}}$ and $\mathbf{s}_{\text{strain}}$ measure the per-element energies and strain, respectively, which determine how much the cloth is stretched.
% \MLE{briefly mention/spell out the four shape descriptors for improved structure, I'll make my pass after that.}
They differ since \(\mathbf{U}\) is usually a squared strain scaled by the material parameters. This implies that \(\mathbf{s}_{\text{target}}\) depends on \(\gamma\), which complicates the computation of the corresponding derivatives. The discrete Fourier transform (DFT) descriptor $\mathbf{s}_{\text{freq}}$ applies the 2D fast Fourier transform (FFT) algorithm to grid vertex positions $\mathbf{p}$ followed by a component-wise magnitude computation, which removes phase information, hence the term \textit{unphased}. This leads to small metric differences when comparing distinct wrinkle phases and larger discrepancies for varying wrinkle amplitude or frequency. 

\begin{figure}
    \centering
    \includegraphics[width=0.85\linewidth,trim=0 0 0 0,clip]{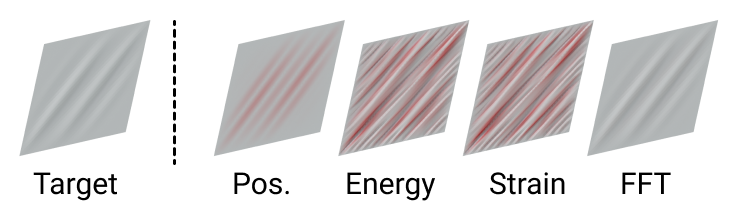}
    \caption{\emph{Shape descriptor evaluation.} %Target cloth is simulated with XPBD (left) followed by 
    Optimization results for each descriptor (right) attempting to match the target reference (left). Euclidean vertex distances to the target are color-coded with a maximum of 1.8 mm.
    % \MLE{mention cloth size earlier (e.g. sec 3.4) to put into perspective}
    %Error metric comparison. Each of the shape descriptors in \secref{sec:shapedescriptors} is used to optimize for a set of parameters using the picture frame and stretch examples. On the left is the desired target generated by XPBD with parameters \(1e5, 2.8e5, 4e4, 0.4, 50\). The following four are the optimization results of each descriptor initialized with parameters $(2e5, 1.8e5, 5e4, 0.3, 10)$. The red color indicates the vertex distance of each result from the target with maximum distance (reddest) is 1.8mm.
    % \TS{What does the red color indicate here? We are trying to recover material parameters for which we know the ground truth values. Perhaps we can add this metric per descriptor in the figure.}}
    }
    \label{fig:error_comparison}
\end{figure}

Our experiments show that the energy and strain descriptors $\mathbf{s}_{\text{energy}}$ and $\mathbf{s}_{\text{strain}}$ work well when bifurcations are symmetric as in \figref{fig:wrinkle}. However, for fine wrinkles as in \figref{fig:error_comparison}, only the frequency descriptor $\mathbf{s}_{\text{freq}}$ performs robustly as we elaborate in the next section.
% \MLE{do we have an example for this statement? energy/strain performing well for single bifurcation and bad for fine wrinkle pattern ambiguities. Otherwise we might need to reformulate this statement}

\section{Evaluation}
\label{sec:evaluation}
In this work, we use the centimeter–gram–second (CGS) system of units for material parameters. %The CGS's equivalent for Newtons (N) is called dyne (dyn) where \(1\text{ N} = 1e5\text{ dyn}\).
While our optimization results in optimal parameters $\gamma$ as defined in \myeqref{eq:params}, we often report the traditional, more intuitive parameters $E_u, E_v, \mu, \nu$ instead of $C_{00}, C_{11}, C_{01},$ and $C_{22}$ as defined in \myeqref{eq:paramsCorrespondence}. In the following sections, we use $\gamma$ to represent parameters in both forms depending on context.
% \MLE{here could be a good place to explain the use of $C_i$,.. and its correspondence to $E_\nu$,.., (and maybe what, for example, $\gamma_\text{cotton}$ means)}

\subsection{Picture Frame Shear}
A popular experiment for cloth is the \textit{picture frame shear}. Here, a square piece of cloth is fixed on all sides onto a rigid ``picture frame'', which is then deformed at the hinges. % where the sides of the frame connect. 
For validation purposes, we hand-pick fictional properties for three synthetic materials, see Table~\ref{tab:synthetic_mat} (top). We call them cotton, denim, and silk as they produce visually similar shapes to these fabrics. 

\begin{table}[b]
\centering
    \centering
    %\footnotesize
    \begin{tabular}{lccccc}
        \toprule
        & $E_u$ & $E_v$ & $\mu$ & $\nu$ & $b$ \\
        \midrule
        \(\gamma_\text{cotton}\)    & $1.0e5$ & $2.8e5$ & $4.0e4$ & $0.4$  & $50$ \\
        \(\gamma_\text{denim}\)     & $5.0e5$ & $7.0e5$ & $2.0e4$ & $0.45$ & $200$\\
        \(\gamma_\text{silk}\)      & $2.0e5$ & $3.0e5$ & $1.5e4$ & $0.35$ & $10$ \\
        \midrule
        \(\gamma_\text{initial,c}\) & $2.0e5$ & $1.8e5$ & $5.0e4$ & $0.3$  & $10$\\
        \midrule
        \(\gamma_\text{freq,c}\)     & $9.98e4$& $2.79e5$& $3.98e4$& $0.40$ & $51.96$ \\
        %$\gamma_{\text{result}} = (99760, 278952, 39816, 0.4018, 51.96)$
        \bottomrule
    \end{tabular}
    \vspace{0.2cm}
    \caption{\emph{Material properties for validation with XPBD.} Hand-picked properties for three synthetic materials (top), perturbed initial conditions for cotton (middle), and the recovered parameter set \(\gamma_\text{freq,c}\) (bottom) with inputs as in \figref{fig:xpbdValidationTargets}. Here, we choose high stiffness values due to finding the simulation to converge to more realistic equilibria while using larger time steps, see Section 3 in the supplemental document for details. }
    % The stiffnesses here are larger than the real-world values in \tabref{tab:realOptimizedParameters} to have XPBD converge to realistic equilibria for larger time steps. See the supplemental document for details.
    %These are used for validating the proposed FFT error metric.
    %\TS{These values are orders of magnitude higher than the optimized real values in Fig~11. Any way to address this a reviewers will probably point this out.}
    \label{tab:synthetic_mat}
\end{table}

%\MLE{%Have this subsection either in evaluation or method.
%Compare different objective functions, e.g. error image from Fig 1 upper right. Emphasise per text that tiny side view is visible on top. Show slide %58. Add cost plots. Add convergence plots for each parameter as you had in the update slides either in this subsection or next one}

\begin{figure*}[p]
\centering
    \includegraphics[width=\linewidth]{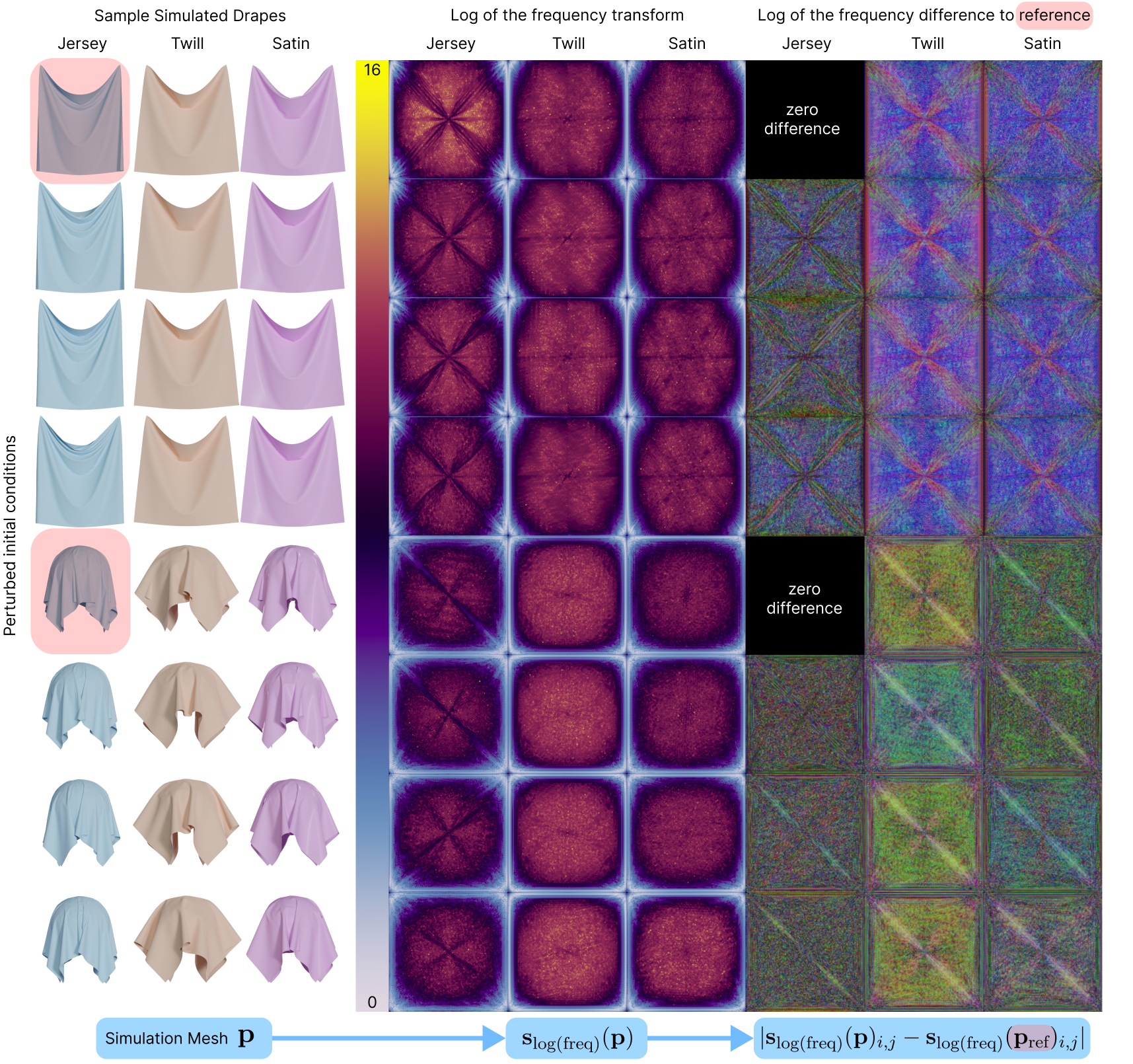}
    \caption{\emph{Frequency descriptor visualization.} In the left section, a square fabric is simulated using 3 different materials in a hanging (top half) and draped on a sphere (bottom half) configurations. Each row corresponds to a different initial condition resulting in slightly different quasi-static equilibria (notice the difference in wrinkling between rows). The middle section shows a visualization of the frequency transform applied to the simulation mesh. The colorization (shown by the color bar) is applied to the natural logarithm of the unphased DFT (denoted by $\mathbf{s}_{\text{log(freq)}}$ to show the emerging patterns characterizing each drape. Notice that similar patterns emerge for the ``Jersey`` material not only between initial conditions but also between the significantly different hanging and sphere drapes.  In the right section, we compute the absolute value of the component-wise difference in $\mathbf{s}_{\text{log(freq)}}$ between each mesh $\mathbf{p}$ and the corresponding highlighted reference $\mathbf{p}_{\text{ref}}$, which generates a zero difference to itself as indicated by the black squares. %Remarkably, ``Twill'' and ``Satin'' meshes differ more from the reference than to other ``Jersey'' simulations as indicated by the brighter colors in the right two columns compared to the ``Jersey'' column.
    }
    \label{fig:fft_demo}
\end{figure*}

To evaluate the effectiveness of the proposed shape descriptors \(\mathbf{s}\), we simulate a picture frame experiment with our simulator and \(\gamma_\text{cotton}\), see target shape in \figref{fig:error_comparison}. Starting with perturbed parameters \(\gamma_\text{initial,c}\) from Table~\ref{tab:synthetic_mat} (middle), we optimize for material parameters with each of the four shape descriptors from \myeqref{eq:shapeDescriptors}. We then run four simulations with the optimized parameter sets, all initialized to the target cloth shape. As demonstrated in \figref{fig:error_comparison}, the optimization using the parameters retrieved through the frequency descriptor \(\mathbf{s}_{\text{freq}}\) converges to the exact target cloth shape while 
%in spite of being more forgiving than the alternative descriptors, which produce distinct cloth configurations 
\(\mathbf{s}_{\text{energy}}\) and \(\mathbf{s}_{\text{strain}}\) produce the largest differences to the target.
% \KW{In Fig. 5, are the distances measured with the different shape descriptors? Or are the shape descriptors used to create the approximations and then the distance is measured as Euclidian distance?}
% \KW{When measuring distances with the FFT, we don't need to reach the exact same cloth shape. Here it sounds like that using FFT, we get the exact same shape nevertheless, even though FFT allows for more freedom. Is this the case? If yes, can we highlight it in the text, I find that a surprisingly great result. If not, can we make that clear?}

\begin{figure}[t]
    \centering
    \includegraphics[width=\linewidth]{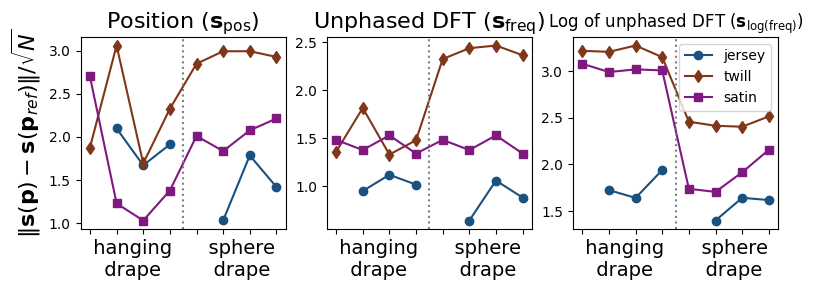}
    \caption{\emph{Comparison of position and frequency based metrics.} The sample drapes from \figref{fig:fft_demo} are compared to the corresponding highlighted reference drape using 3 different shape descriptors. The norm of the shape descriptor differences are plotted. The reference configurations are omitted since they would produce a zero when compared with themselves (indicated by the gaps in the ``jersey'' plots).
    The leftmost plot demonstrates that the position norm is unable to distinguish between different materials. In contrast, the unphased frequency descriptor (middle) shows better separation between different materials. The natural logarithm of the unphased frequency (right) further separates the different materials. Although this effect is not guaranteed we notice that it is valuable in practice.
    }
    \label{fig:normalizedMetricDifferences}
\end{figure}

\subsection{Drape Frequency Analysis}
To illustrate how a frequency-based metric is useful for distinguishing fabrics of different materials we visualize the unphased DFT of cloth simulated with 3 different material properties with 4 perturbed initial conditions in 2 configurations in~\figref{fig:fft_demo}. We compute the unphased DFT and take the logarithm of the result for each component of each vertex. By taking the log of each component and taking the 2-norm of the 3 components reveals distinct patterns in each of the distinct materials. This exercise reveals that in frequency space, different or perturbed drapes of the same material are more similar than drapes of different materials, for which the fabric wrinkles differently, see~\figref{fig:normalizedMetricDifferences}. We also visualize the differences of each drape to the highlighted reference of the corresponding drape to show that indeed the same material exhibits smaller (darker in the ``Jersey'' column in the rightmost section of~\figref{fig:fft_demo}) differences in frequency space than drapes for different materials. This example motivates the use of frequency based norms for measuring differences in different fabrics. While in our loss we use frequency values without the logarithm, it remains a promising research direction to discover better metrics.

\subsection{Validation}

%\todo{Can we compare to material values optimized through the FAB pipeline?}

% \MLE{Copied from intro:}
% \MLE{Proof of correct parameter estimation (= the method is working as we claim): 
% evaluation on synthetic data: show that algorithm is able to determine correct material parameters of 1-3 different fabrics based on different (?) input cloth states (different force application if available / if makes sense to you) }
%EL: Looks like I never ran the same bend-then-stretch scenario as i did for real and synthetic data. All I have is picture frame shear with stretch, which, arguably should make it harder for the optimizer, but still pretty unsatisfactory. Maybe the reviewers will ask for better proof and we can rerun the xpbd validation with the same scenarios.
% EL TODO: fix formatting of the inset figure here (can be smaller and smaller margins). 

To analyze the accuracy of our parameter estimation pipeline, we generate four synthetic cotton simulations with our simulator: two picture frame deformations and two stretch experiments
\begin{wrapfigure}{r}{2cm}
    \centering
    \includegraphics[width=\linewidth]{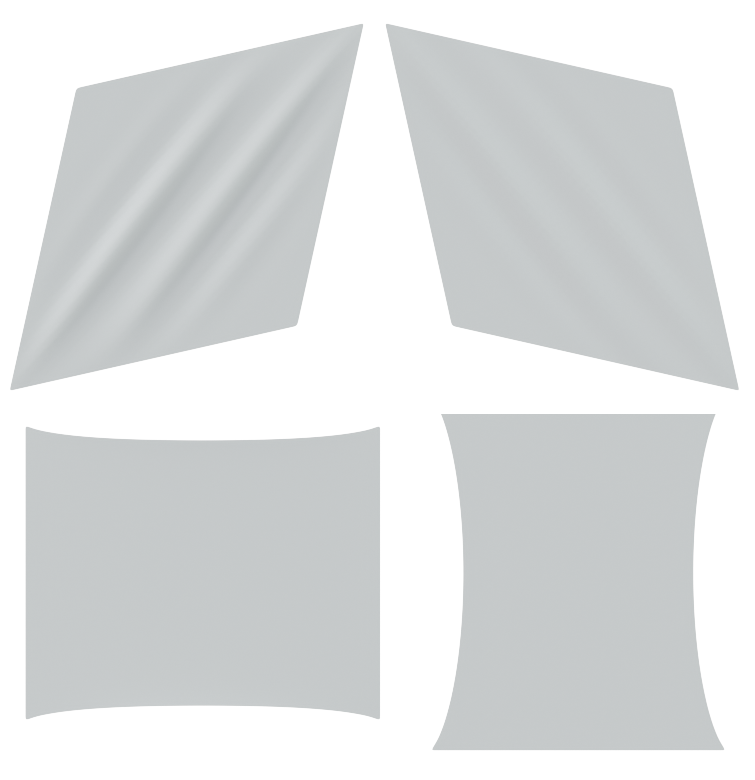}
    \caption{\emph{Validation targets.}\label{fig:xpbdValidationTargets}}
\end{wrapfigure}
as visualized in \figref{fig:xpbdValidationTargets}. 
To allow for precise comparison of the retrieved and target parameter sets, we use the same simulator as we employ in the optimization. Starting with the same perturbed parameter set \(\gamma_\text{initial,c}\) as in the previous section, our optimizer recovers the target parameters very well after 13 iterations as shown in Table~\ref{tab:synthetic_mat} (bottom) -- although only four target cloth shapes are used. Remarkably, the bending parameter is recovered within a $4\%$ error in spite of there being no explicit bending examples in the target set. The membrane parameters are recovered within $0.5\%$ error. 

\begin{figure}[ht]
    \centering
    \includegraphics[width=\linewidth,trim=0 0 0 0, clip]{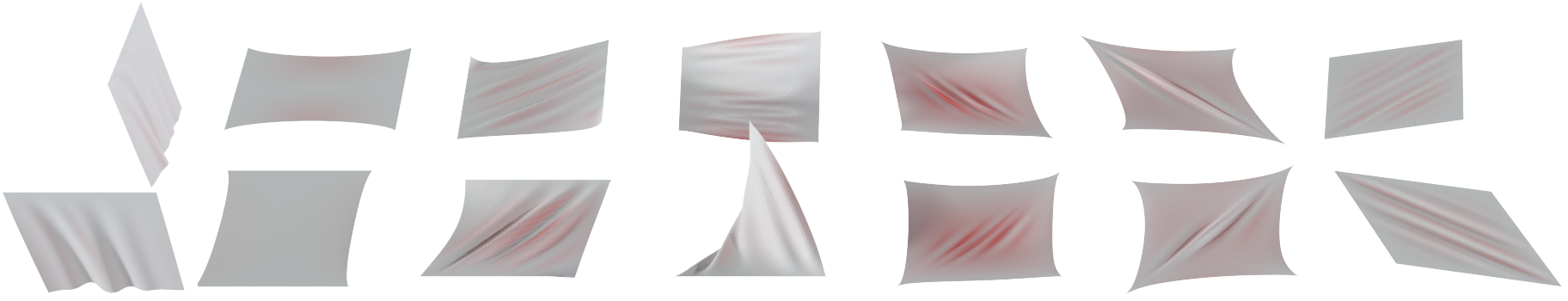}
    \caption{\emph{Re-simulation of various cloth configurations with $\gamma_{\text{freq,c}}$ estimated from the targets using our optimization method with $\gamma_{\text{cotton}}$}. The vertex displacement error is color-coded in red, with a small maximum error of 0.28 mm. The second and last columns are targets while the rest are unseen by the optimization.}
    \label{fig:xpbdValidation}
\end{figure}

In \figref{fig:xpbdValidation} we further compare the shapes of unseen cloth configurations re-simulated with our retrieved parameter set \(\gamma_\text{freq,c}\) to target simulations created with \(\gamma_\text{cotton}\).
The error in the recovered parameters causes a difference in vertex positions below 0.28 mm, which is largely imperceptible. 

\section{Results}

%All experiments are run using centimetre–gram–second (CGS) system of units. The CGS equivalent for Newtons (N) is called dyne (dyn) where 1 N equals $1e5$ dyn.

% Removing this result in favour of fig 5 which gets a much better estimate.
% \begin{figure}[h]
%     \centering
%     \includegraphics[width=0.45\textwidth]{images/xpbdFFTCottonBendAndStretch.png}
%     \caption{Our optimization accurately recovers the simulation generated by our internal XPBD solver (maximum 0.5mm error vertex displacement error). Note that we don't expect the parameters to be recovered exactly since the FFT objective deliberately ignores exact vertex position information.
%     We estimated the parameters using only bending, pulling corners and stretching scenarios.
%     Initial parameters: 110000 270000 41000 0.41 51 \\
%     Target parameters: 100000 280000 40000 0.40 50 \\
%     Best result: 101630 286186 39524.7 0.39261 48.6369
%     \EL{Trying with a larger perturbation since the result is not even that far from the perturbation.}
%     This shows that the method works.}
%     \label{fig:xpbdValidation}
% \end{figure}

\begin{figure}[b]
    \centering
    % TODO: Switch to cotton
    \includegraphics[width=0.98\linewidth,trim=0 0 0 0 0, clip]{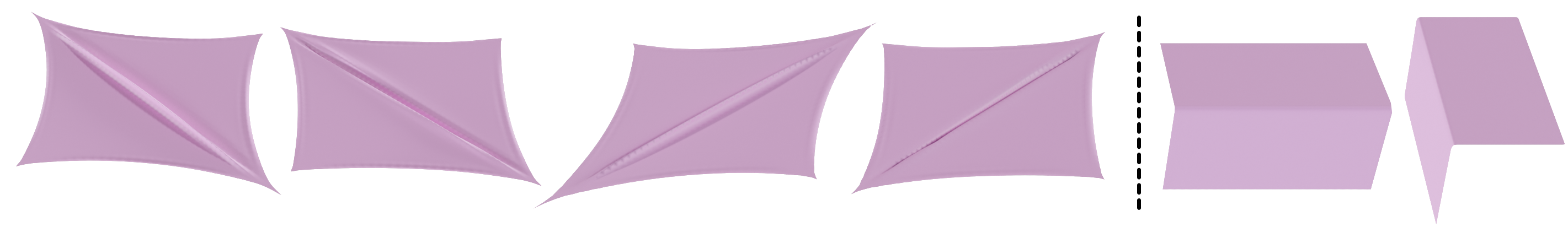}
    \caption{\emph{Synthetic silk targets generated with Houdini,} used for estimating membrane stiffnesses (left) and bending stiffnesses (right).
    %\TS{is it silk or cotton? Both are mentioned here}.
    }
    \label{fig:houdiniTargets}
\end{figure}

\paragraph*{Synthetic Experiments.}
We show that our method is capable of reproducing cloth simulations from third-party software. First, we generate a set of square swatch targets in Houdini \cite{sidefx19} with different material presets; the targets for silk are depicted in~\figref{fig:houdiniTargets}. 
Then, we estimate the bending parameter $b$ using the bending scenarios of \figref{fig:houdiniTargets} (right). The membrane parameters $(C_{00}, C_{11}, C_{01}, C_{22})$ are estimated together in a separate pass by pulling the corners of the suspended cloth swatch as in \figref{fig:houdiniTargets} (left).
This process is repeated thrice to ensure that the coupling between bending and membrane stiffnesses is not lost. 
Furthermore, to avoid getting stuck in local minima, we run the same optimization from three different randomized starting points.

\begin{figure}[t]
    \centering
    \includegraphics[width=0.45\textwidth,trim=0 400 0 300,clip]{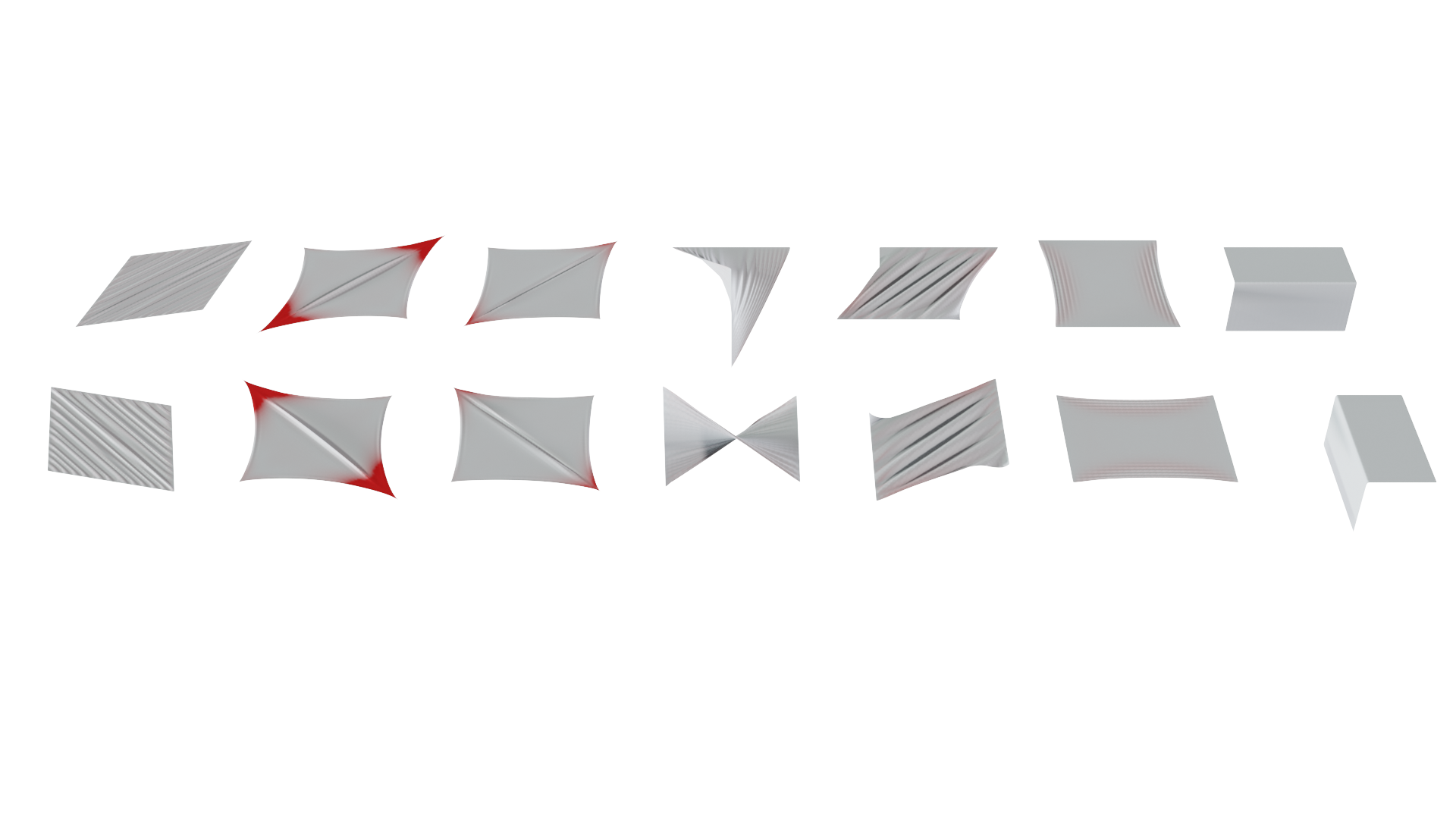}
    % %\includegraphics[width=0.33\linewidth,trim=100 0 100 0,clip]{images/CloseupSilkStretchWeftTarget.png}%
    % %\includegraphics[width=0.33\linewidth,trim=100 0 100 0,clip]{images/CloseupSilkStretchWeftInitial.png}%
    % %\includegraphics[width=0.33\linewidth,trim=100 0 100 0,clip]{images/CloseupSilkStretchWeftBest.png}
    % \begin{tabular*}{\linewidth}{ccc}
    %     \centering
    %     \includegraphics[width=0.3\linewidth,trim=100 0 500 100,clip]{images/CloseupSilkStretchWeftTarget.png} &
    %     \includegraphics[width=0.3\linewidth,trim=100 0 500 100,clip]{images/CloseupSilkStretchWeftInitial.png} &
    %     \includegraphics[width=0.3\linewidth,trim=100 0 500 100,clip]{images/CloseupSilkStretchWeftBest.png} \\
    %      Target & Initial & Optimized
    % \end{tabular*} 
    \caption{\emph{Re-simulation with the estimation of Houdini's silk preset.} The vertex displacement error is color-coded in red with a maximum set to 2 mm. In the second column, errors are clustered on the large stretch areas with a maximum of 6 mm while it is below 2.4 mm in the remaining columns.
    }
    \label{fig:syntheticResults}
\end{figure}

This decoupling technique produces a lower objective, i.e., a better fit, compared to optimizing all parameters at once. In the future, we aim to further eliminate the need for separate passes for bending and membrane parameters to produce a more streamlined pipeline.
%, see \secref{sec:limitations} for details
% \MLE{this is not contradicting to 3.4.2, right?}
The comparison to Houdini silk targets is illustrated in \figref{fig:syntheticResults}. For all deformations apart from the second column, the local vertex displacement is below 2.4 mm. In the second column, the error is clustered at large strain regions, which indicates that our StVK model deviates from Houdini's material model for large strain deformations. 

\begin{figure}[b]
    \centering
    \includegraphics[width=0.7\linewidth,trim=0 0 0 0,clip]{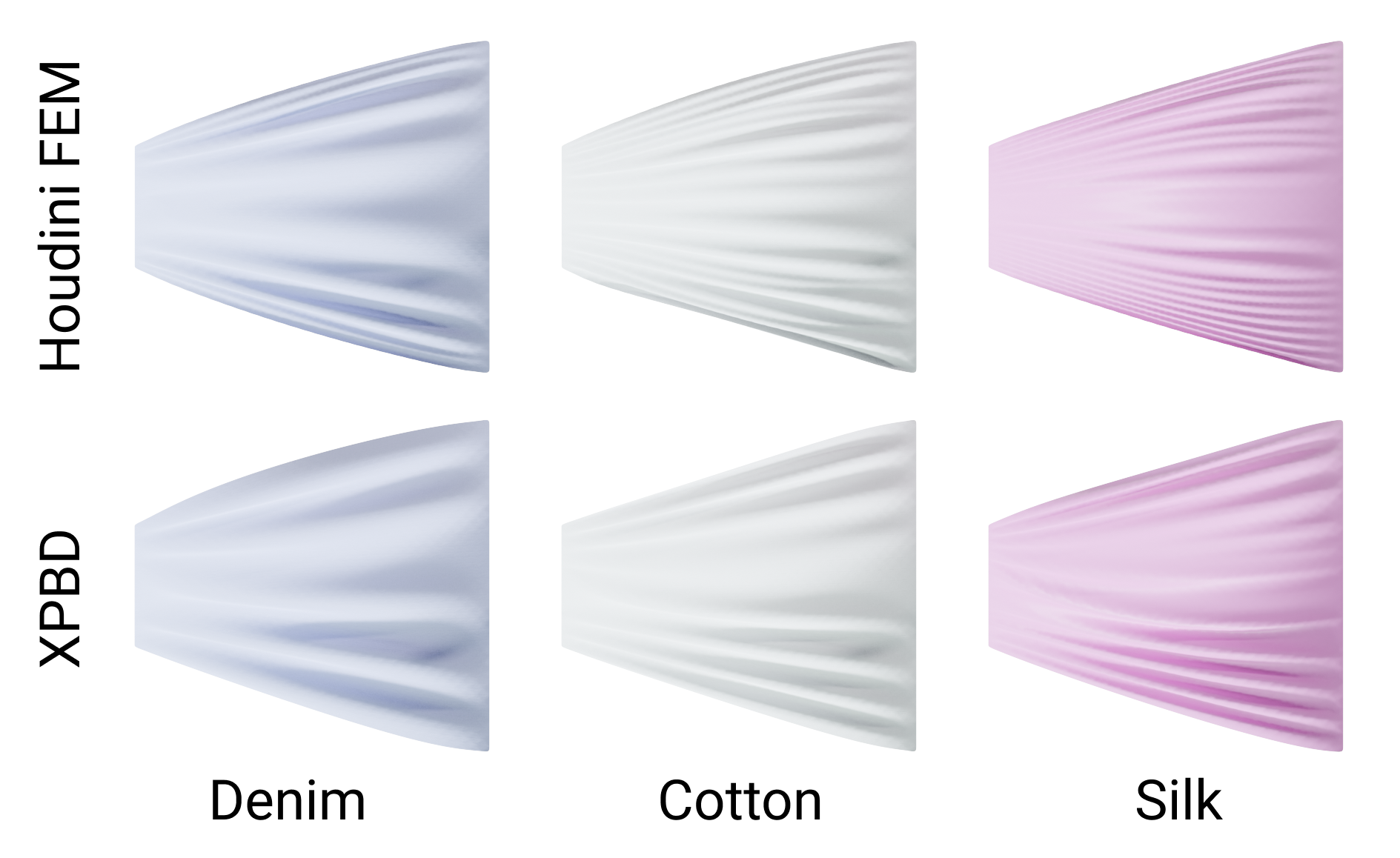}
    \caption{\emph{Aesthetic evaluation}. By twisting a piece of cloth, unseen by optimization, we demonstrate that distinct wrinkle patterns generated by Houdini's FEM simulator are reproduced using our simulator with optimized material parameters. 
    %Our wrinkles being a bit coarser. %The wrinkles on the silk are not as fine as Houdini's target, possibly because our bending parameter is clamped below 0.1 during optimization to avoid producing unrealistically low bending stiffness.
    % \MLE{Would be great to use same initial conditions instead of target vertex positions}
    \label{fig:synthLook}}
\end{figure}

The retrieved parameters are listed in \tabref{tab:syntheticOptimizedParams}. 
% \MLE{can we infer any insight from these parameters? probably not, because we don't know Houdini's material model and corresponding parameters? maybe add another visual comparison for cotton (as for silk) -- or just color-code the cotton results and refer to that as example target experiments?}
% \MLE{the silk example isn't mentioned here yet}
By optimizing for these parameters, we avoid manual hand-picking of material parameters as in Section \ref{sec:evaluation}, but instead, are able to reproduce the cloth behavior of the physically more accurate and computationally more expensive Houdini FEM simulator as demonstrated in \figref{fig:synthLook}. Here, we re-simulate the Houdini experiment with our simulator by using the same initial and boundary conditions as in Houdini. We thereby demonstrate that our pipeline is capable of effectively reproducing the aesthetics of different cloth materials generated by third-party simulators. 
\begin{table}[b]
    \centering
    \footnotesize
    \begin{tabular}{r|ccccc}
        \toprule
         Material ($\rho$ in g/cm${}^2$) & $E_{u}$ & $E_{v}$ & $\mu$ & $\nu$ & $b$
        %  & $r$
         \\
         \midrule
         Denim (0.0324) & 34182 & 33735 & 5592 & 0.2609 & 11.0 \\%& 8.83  \\
         Cotton (0.0224) & 30823 & 30266 & 4971 & 0.3058 & 1.19 \\%& 13.9\\
         Silk (0.0187) & 7492 & 7436 & 1193 & 0.03543 & 0.100 \\%& 18.6 \\
         \bottomrule
    \end{tabular}
    \vspace{0.2cm}
    \caption{\emph{Estimation of material parameters for Houdini's FEM cloth solver.}}
    % In the last column, $r = \|\mathbf{r}_{\text{sim}} - \mathbf{r}_{\text{target}}\|^2$ reports the squared residual norm using \emph{all} of the configurations as shown in \figref{fig:syntheticResults}, including those unseen by the optimization.
    \label{tab:syntheticOptimizedParams}
\end{table}
% effective parameters
%By matching material parameters to a more physically accurate cloth simulation like Houdini's FEM, we can find effective material parameters in our XPBD simulator to produce similar behavior without paying for the additional accuracy of the more expensive simulation.
% \TS{\figref{fig:synthLook} just shows results from Houdini right? How does this relate to our pipeline?} \MLE{Yes we need to show our simulator's results with the parameters in Table 2, best would be to have it side by side with Houdini's results to prove that we match the look}

% \MLE{TODO describe Fig. 11 and its purpose (draw conclusion)}

% \MLE{Adjust when section is written:
% This demonstrates that our method can be used for reproducing the cloth behavior of other third-party tools without the need to hand-pick material parameters to match a certain look. }

% \EL{TODO: Add comparison against the sequence from \cite{miguel12}. Our bend->stretch strategy for optimization seems to give better results. Just showing best outcomes from both should be convincing enough}

% EL: We have not done any of this yet, omitting for sig submission:
% \paragraph{Validation.}
% \MLE{copied from intro:}
% results for real-world captures: obtain same parameters from different data set \TS{Show 2 different captures of same material under same external forces and show our method still works.} \MLE{also show two captures with different force application and same material leading to same material parameters?}

\begin{figure}[h]
    \centering
    \includegraphics[width=\linewidth,trim=50 500 50 350,clip]{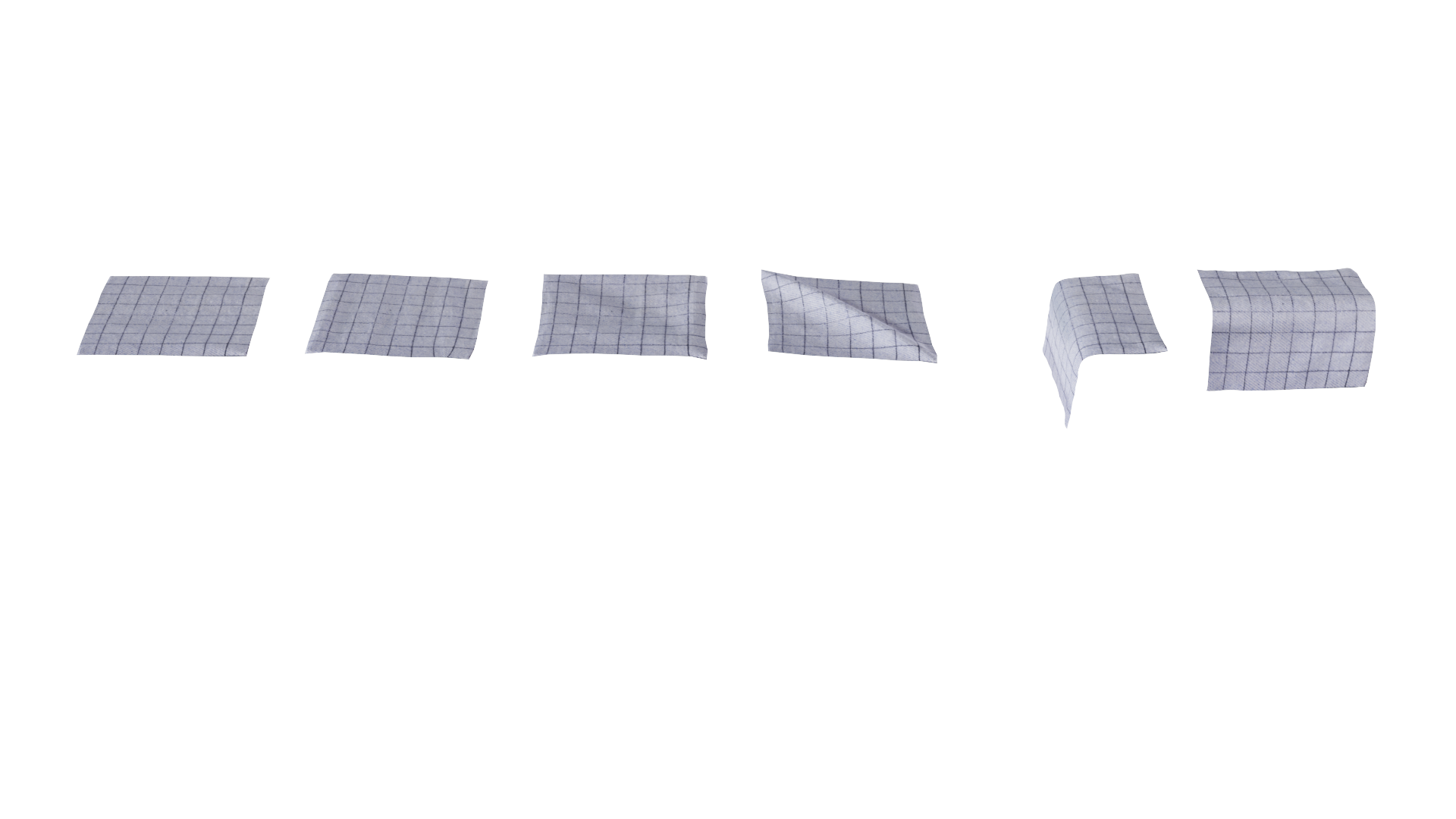}
    \includegraphics[width=\linewidth,trim=50 500 50 350,clip]{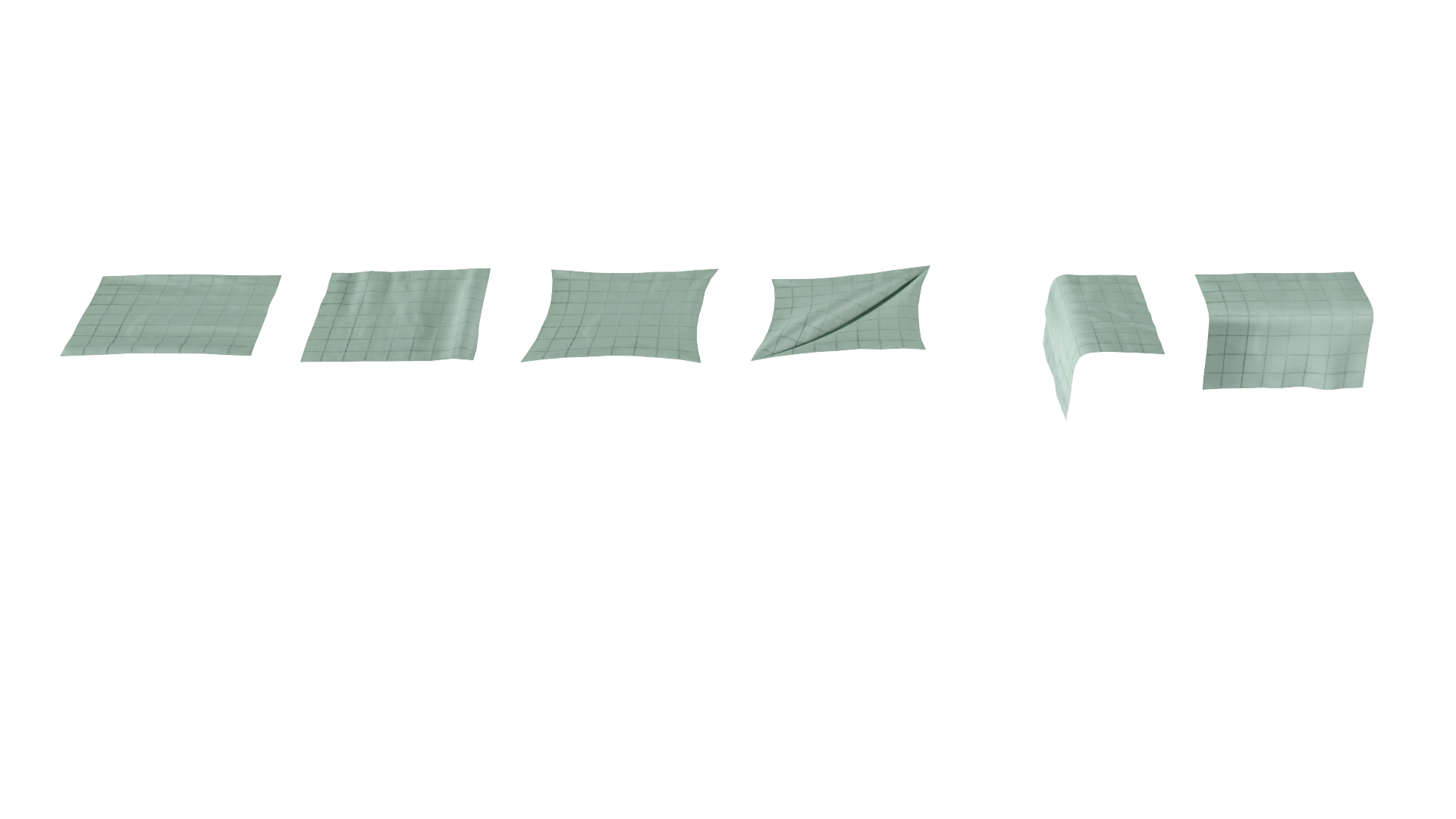}
    \includegraphics[width=\linewidth,trim=50 500 50 350,clip]{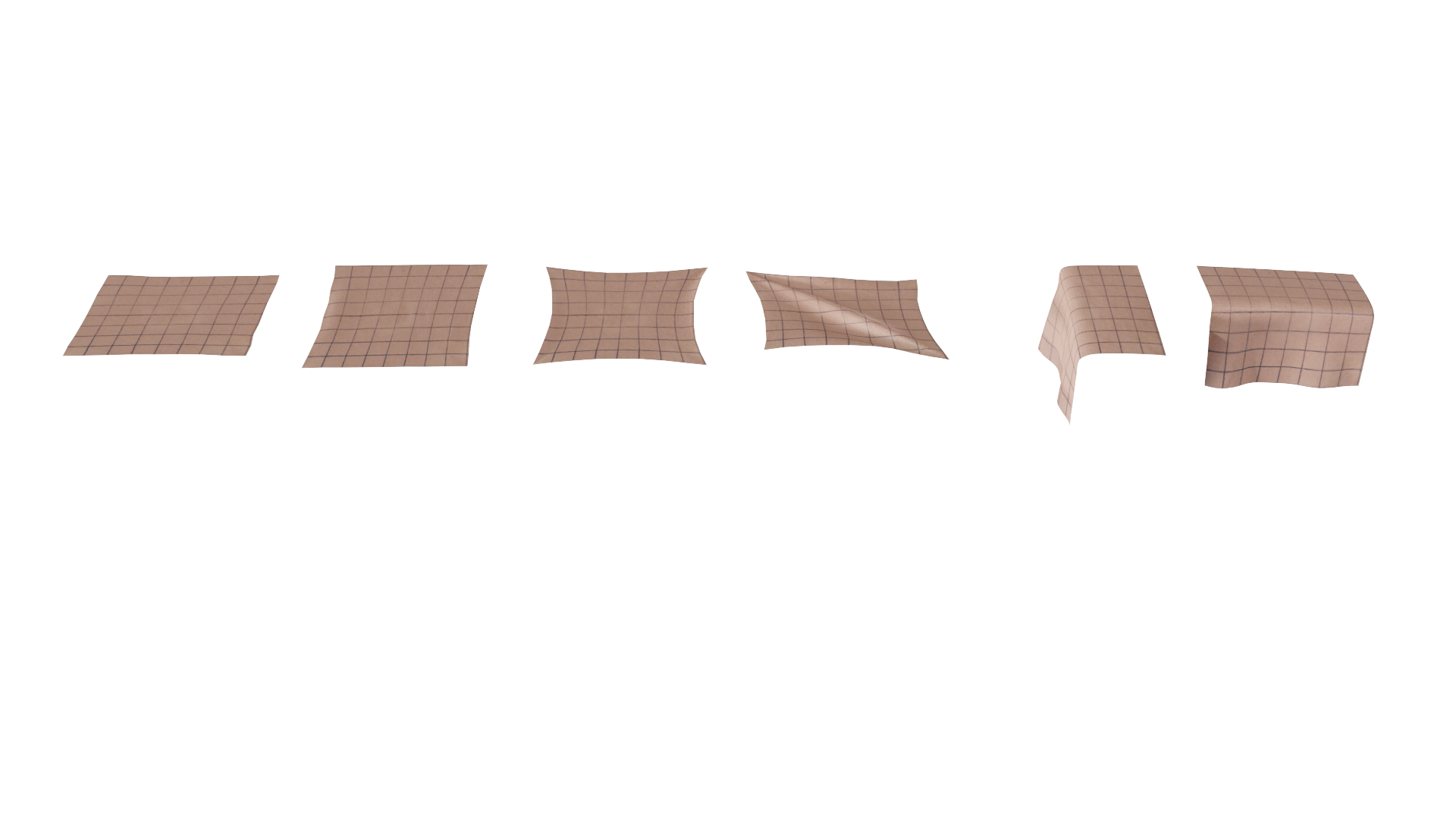}
    \caption{\emph{Real-world targets.} A piece of denim (top), cotton (middle), and polyester (bottom) is deformed in various configurations. The target meshes are generated through template registration as described in \secref{sec:registration}.}
    \label{fig:realTargets}
\end{figure}

\paragraph*{Real-world Experiments.} With real-world captures of cloth subject to various force applications and deformations, we generate six targets each for three different materials as shown in \figref{fig:realTargets} using a 9.6 cm cloth swatch.
The first two targets are generated by applying a 200 g weight on opposing sides of the cloth using wide paper clamps as shown in \figref{fig:captureSystem} (right). The third target is generated by applying a 200 g weight on all corners while the fourth is created by applying 20 g at two diagonally opposite corners and 100 g on the remaining corners. The last two targets are generated by letting the cloth drape from a horizontal ledge.

\begin{figure}[h]
    \centering
    \includegraphics[width=\linewidth,trim=50 500 50 350,clip]{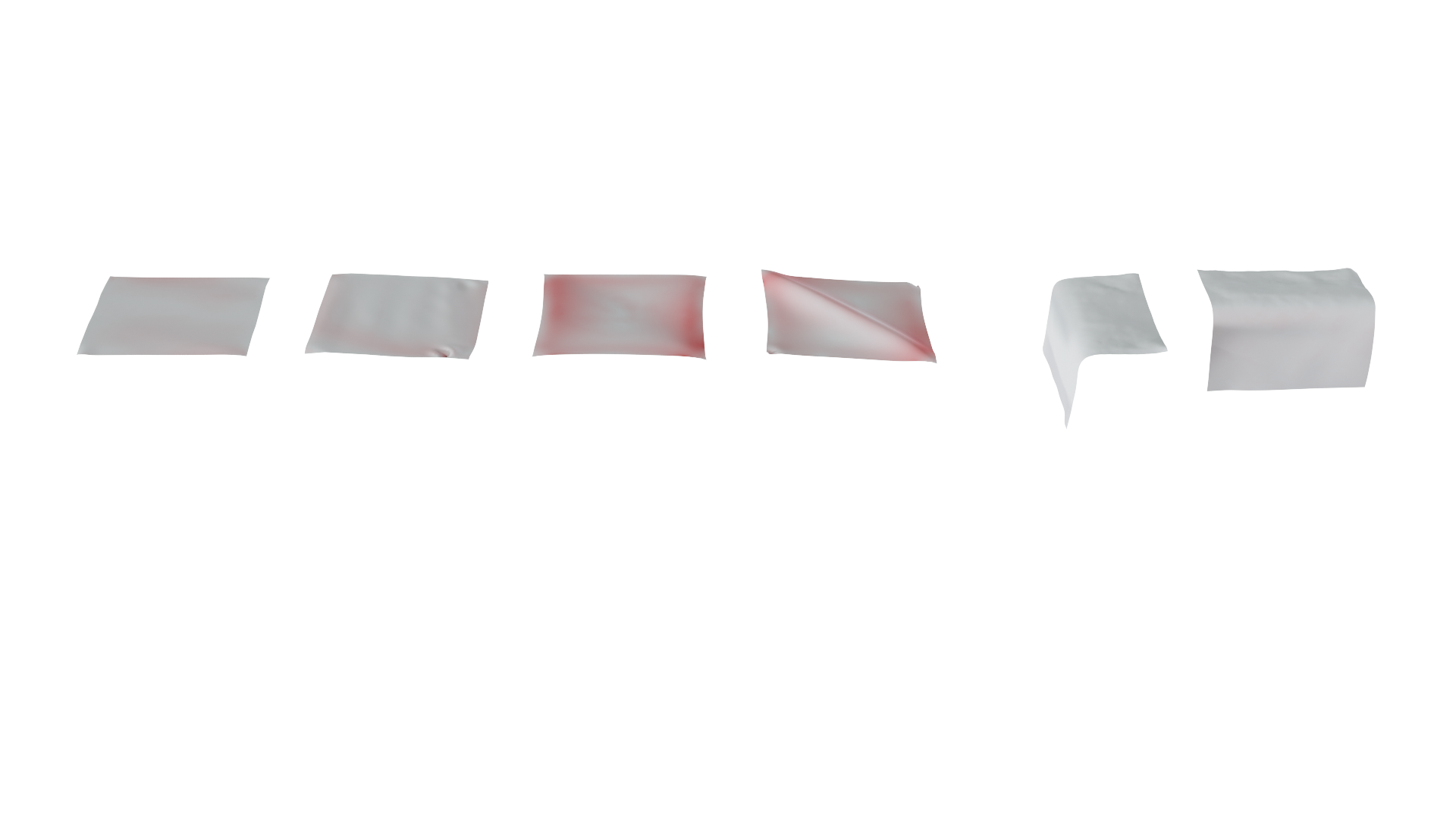}
    \includegraphics[width=\linewidth,trim=50 500 50 350,clip]{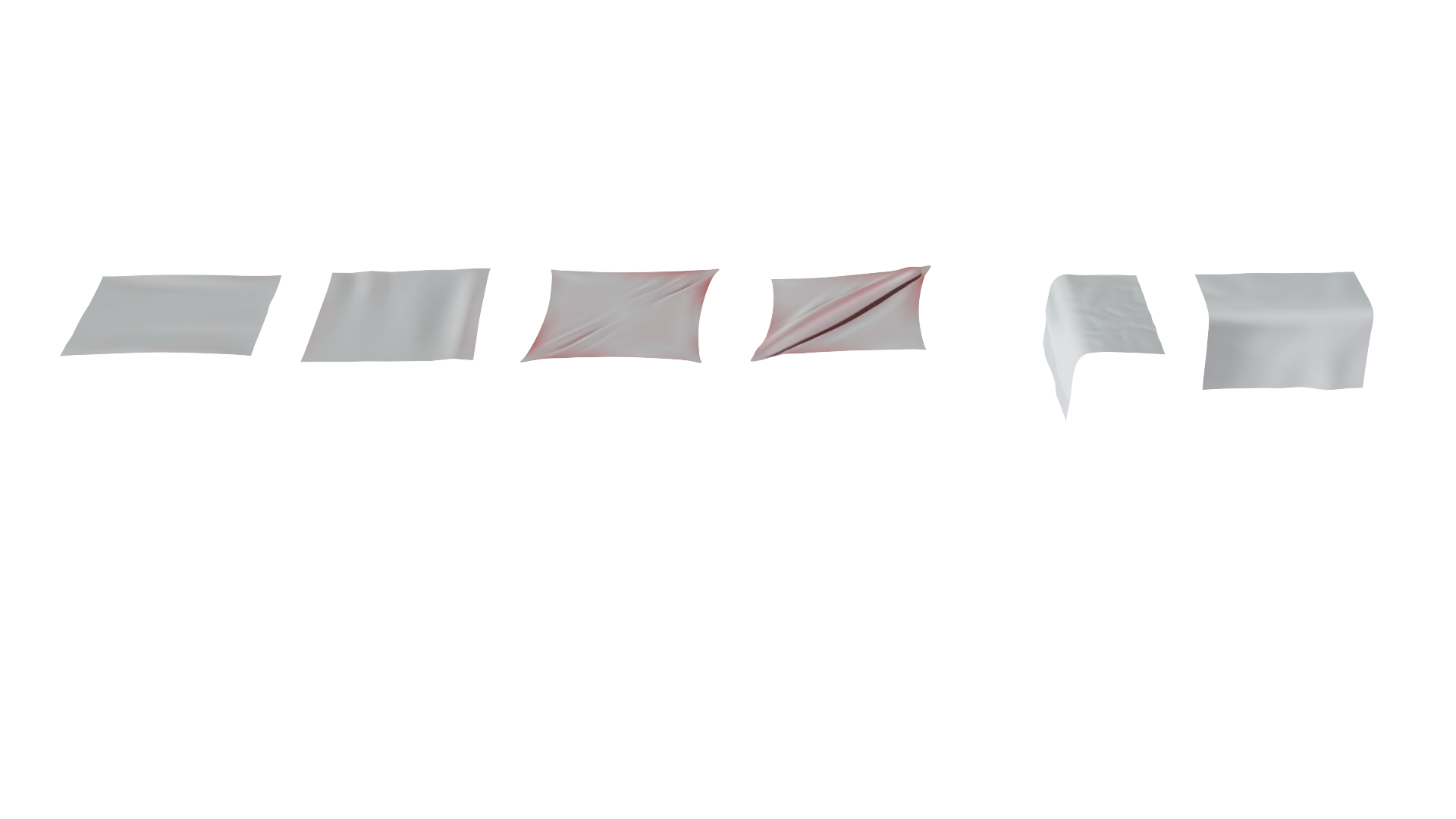}
    \includegraphics[width=\linewidth,trim=50 500 50 350,clip]{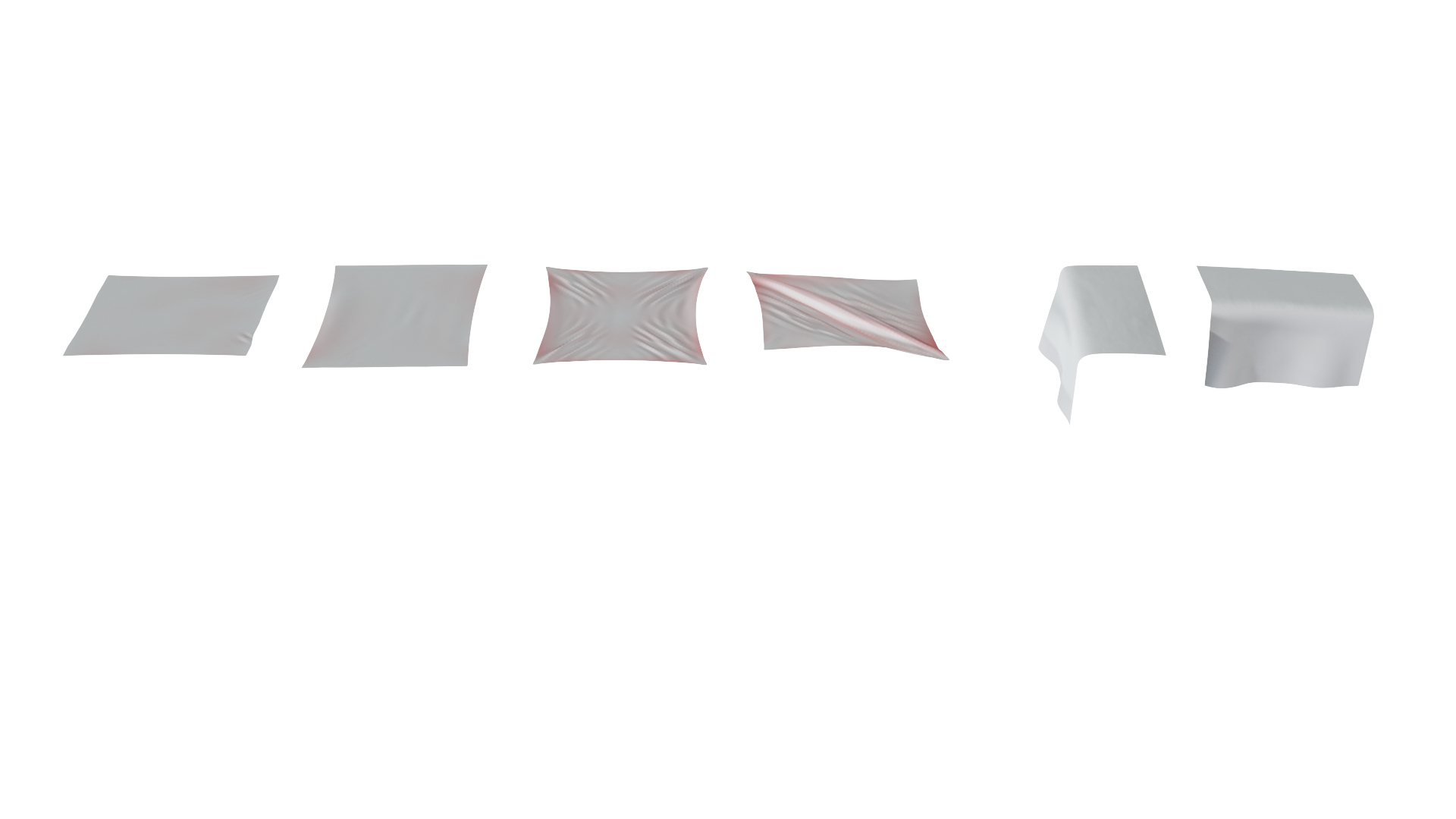}
    \caption{\emph{Real-world results.} Denim (top), cotton (middle), and polyester (bottom) cloth is simulated with our estimated parameters. The vertex position error to the targets in \figref{fig:realTargets} is colored in red with a maximum of 3.7 mm.}
    % The resulting parameters for denim were $(4212, 23540, 6584, 0.42, 24)$ and for polyester $(1219, 4252, 5941, 0.19, 6)$. Unsurprisingly, the polyester is significantly less stiff.}
    \label{fig:realResults}
\end{figure}

We optimize for the parameters of each material, which are shown in \tabref{tab:realOptimizedParameters}. Then, we re-simulate the target with the estimated parameters by setting the initial vertex positions to the target shapes, see \figref{fig:realResults}.
%Our results demonstrate that our method is able to closely match the captured targets within 3.7 mm of maximum vertex displacement. 

Our optimization correctly determines higher Young's moduli for a stiff material like denim and lower for the softer polyester material, as well as a larger bending stiffness in denim compared to cotton and polyester.
\begin{table}[h]
    \centering
    \footnotesize
    \begin{tabular}{r|ccccc}
        \toprule
         Material ($\rho$ in g/cm${}^2$)& $E_{u}$ & $E_{v}$ & $\mu$ & $\nu$ & $b$
        %  & $r$
         \\
         \midrule
         Denim (0.0324) & 3793 & 20590 & 6968 & 0.4286 & 28.5 \\%& 3.95 \\
         Cotton (0.0224) & 1840 & 2019 & 6538 & 0.4308 & 8.99 \\%& 1.31 \\
         Polyester (0.0187) & 1028 & 3271 & 9171 & 0.2731 & 2.35 \\%& 4.32 \\
         \bottomrule
    \end{tabular}
    \vspace{0.2cm}
    \caption{
    \emph{Estimation of material parameters for real-world targets.}
    %\emph{Optimized parameters for real-world examples.} Resulting optimized parameters for our StVK model using real-world targets. 
    % In the last column, $ r = \|\mathbf{r}_{\text{sim}} - \mathbf{r}_{\text{target}}\|^2$ reports the squared residual norm for all membrane and bending targets together.
    }
    %EL: Woot as expected polyester got the lowest bending parameter
    \label{tab:realOptimizedParameters}
\end{table}

Note that the stiffness parameters in \tabref{tab:realOptimizedParameters} are lower than in \tabref{tab:syntheticOptimizedParams}, since the forces applied on the boundary are smaller in the real-world case, e.g., 0.2 Mdyn vs. 3 Mdyn, but leads to a similar amount of strain.
Since our method aims at small strain deformations, we expect to see lower stiffness estimates.  Furthermore, for lower stiffnesses, longer simulation times may be needed for the configuration to deform sufficiently to produce a signal in the proposed losses. We propose future directions to address this in \secref{sec:limitations}.
% We plan to extend our analysis to more sophisticated large strain models, and experiment with larger forces for captured data in the future.
%in synthetic in the cotton stretch example) 
% This indicates that Houdini presets may be producing stiffer materials than in reality.
% \todo{In either case the optimized Young's moduli are substantially lower than theoretical values. 
% This discrepancy emphasizes the importance of automatic parameter optimization, since theoretical stiffness values cannot be applied directly in cloth simulation engines used for animation.}
% \todo{Mention small strain assumption in our work.}

%\TS{Show working optimization results for different cloth materials and subjective validation of a full garment simulated with the optimized material parameters. Compare with real piece of clothing on a mannequin?}

\begin{figure}[h]
    \centering
    \includegraphics[width=0.49\textwidth]{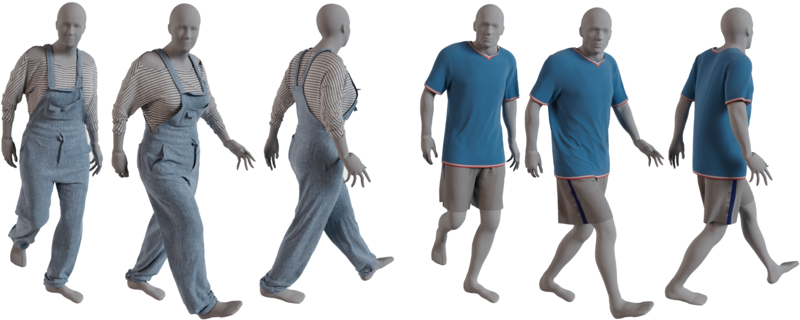}
    \caption{\emph{Outfit simulation:} a denim overall and cotton top (left) and a polyester soccer outfit (right). }%\MLE{I guess we won't have time for this but it would be awesome to show fail examples, e.g., the overall as polyester and the soccer as denim; this would demonstrate how critical material parameters are for the perceived "correct look". }
    \label{fig:simFullGarment}
\end{figure}

To further validate that our results produce the desired look, we simulate full outfits using the estimated parameters in \figref{fig:simFullGarment}.

% \TS{TO DO: Add experiments we want to show and work towards creating these.}

%\paragraph{Show that the system is robust against wrinkles.}
%\TS{Ablation study of the different terms in the loss function}

% \TS{Experiment showing the bifurcation behavior of cloth making positional loss a poor choice. E.g. the ball dropping on swatch example.}
% EL: Moved to Evaluation section "Choice of Objective Function"

% \paragraph{Show that the system is more flexible than previous works.}
% \EL{Show that the same/similar parameters can be estimated using a different data set.}
%EL: We dont have good examples of this unfortunately.

% Postponed: no time for this for sig.
% \paragraph{Compare with currently adopted method(s).}
%
% \EL{
% Compare against Argus/Arcsim \cite{wang2011data} measured materials
% \begin{itemize}
%     \item Test how our StVK model compares to piecewise linear in \cite{wang2011data}.
%     \item If the comparison is not favourable, then address limitations in material model (new material model).
% \end{itemize}
% }

%\input{comparison.tex}

\section{Limitations and Future Work} \label{sec:limitations}

We presented a novel approach for estimating elasticity parameters for cloth simulation using a frequency-based metric that extracts wrinkling similarity between meshes corresponding to simulated fabric.
Our approach is a step forward to simplifying cloth captures and optimization pipelines used for inverse design and parameter estimation.
Despite improving robustness and simplicity over prior work, several limitations remain. For instance, the accuracy of our template registration technique is limited to the resolution of the stamped pattern. %Furthermore, the landmark vertex selection on the scanned mesh is an arduous process. We plan to improve on these with a more sophisticated correspondence method.
Additionally, as is common with nonlinear optimization, our system is sensitive to objective scaling even with our frequency-based metric, and heuristics are required to establish a reasonable scaling between the shape descriptor metric and the force metric. %Furthermore different deformation scenarios can systematically exhibit larger errors than others, which biases the objective. Similarly, sensitivity to initial parameters is another issue we aim to improve upon. Currently, to alleviate this, we run each of the experiments with 3 trials with randomized initial conditions. However, our hand-picked starting point worked best in all of our experiments. It would be valuable to further investigate the local minima in the parameter space.

%Although several simulation methods would be adequate to optimize for material parameters, we opted to leverage XPBD as part of the novel contribution. In this paper, we use a multi-threaded CPU implementation. 
% As part of future work,
%Even though XPBD can be run in real-time, to obtained fully converged results can take longer.
In the future, we intend to extend our research to different material models such as the piece-wise linear elastic model \cite{wang2011data} or spline model \cite{clyde17} to better match experimental data. Additionally, research into perceptual or feature-based cloth quality metrics offers the potential to further embrace the inherent coupling of cloth material parameters. 
%\KW{I don't get the last sentence. The coupling of cloth material parameters? Or having different materials? And do you mean here that we replace the stamped grid with another method, and that will improve the registration?}
% Additional, it will be interesting to further look into motion sequences and dynamic material parameters. We anticipate this to be a more informed approach to parameter estimation rather than using a few static scans, since this will provide substantially more data to work. Capturing and modeling of cloth hysteresis effects also remains as future work.
% \EL{TODO: (Maybe?) mention that because the sims ran for a limited time, some optimization \emph{may} bias lower stiffness values, since those would cause the simulation to deviate slower from the target and thus end up in a closer approximation. To analyse this potential bias we propose to repeat the same optimization over different simulation times, to find the appropriate time. It's worth noting that the stiffness cannot be too low since all sims are subject to gravity.}

Our method assumes a constant simulation time to reach equilibrium. Although, in practice we find that convergence to equilibrium is not strictly necessary for achieving reasonable results, it should improve the performance of the frequency-based metric, which may be sensitive to intermediate wrinkled states (See the polyester example on the bottom 3rd column of \figref{fig:realResults}, which shows high frequency wrinkles of a non-equilibrium state).  %Furthermore, constant simulation times may pessimize stiffness estimation since at lower stiffnesses, simulation takes more time to reach an equilibrium state.  While time to equilibrium is generally short close to the optimum, we can expect it to increase when initial parameters are chosen poorly.  This can be improved by initializing each simulation using last known best iteration, which ought to improve performance and thus allow for longer simulation times.
Furthermore, we aim to implement the method on GPUs to optimally leverage available compute resources and to greatly speed up to optimization time. This will, in turn, enable longer simulation times to allow challenging configurations to reach equilibrium states.

\section{Conclusion}

We have presented a novel, simple method to capture, register, and optimize for cloth material parameters using a differentiable simulation engine. Our proposed pipeline consists of three separate stages that can be improved upon individually. We proposed a novel metric that enables us to capture coupled cloth shearing, stretching, and bending effects, which in turn allows for more efficient and easier capture setups. Although we demonstrate our method using a compliant constraint simulation method, the proposed method would work with other simulation frameworks. Optimized material parameters can be reused with different simulation techniques that employ the same material model.

% \begin{acks}
%\section{Acknowledgements}
%We would like to thank Michelle Hill and Arkell Rasiah for their help with the cloth captures.
% \end{acks}

% \bibliographystyle{ACM-Reference-Format}
\bibliographystyle{acm/ACM-Reference-Format}
\bibliography{bibliography}

%\appendix

\end{document}